\documentclass[12pt]{article}

\usepackage{extsizes}
\usepackage{graphicx}
\usepackage{amsmath,amsthm}
\usepackage{amssymb,latexsym}
\usepackage{enumerate}
\usepackage{epigraph}
\usepackage[authoryear]{natbib}
\usepackage{mathrsfs}
\usepackage{verbatim}
\usepackage{mathabx}
\usepackage{relsize}
\usepackage{mathtools}
\usepackage{accents}
\usepackage{bm}
\usepackage[titletoc,title]{appendix}
\usepackage{fancyhdr}
\usepackage[shortlabels]{enumitem}
\usepackage{tikz}
\usepackage{lipsum}
\usepackage[toc]{glossaries}
\usepackage{makeidx}
\usepackage{url}
\usepackage{imakeidx}
\makeglossaries
\makeindex
\indexsetup{othercode=\small}

\theoremstyle{plain}

\theoremstyle{definition}

\theoremstyle{remark}

\newglossaryentry{1/N.rule}
{
  name={$1/N$ rule},
  description={A probability rule that assigns to each of $N$ possibilities a probability of $1/N$ when very little is known about them. Same as \gls{law.of.equal.ignorance}}
}

\newglossaryentry{911}
{
  name={$9$/$11$},
  description={An abbreviation for the date of the World Trade Center attacks of September 11, 2001. Apart from its political and historical significance, 9/11 had a significant impact on the U.S. economy}
}

\newglossaryentry{accuracy}
{
  name={accuracy},
  description={In the theory of statistical estimation, an estimator is accurate if the expected value of its difference from the parameter is small, otherwise, inaccurate}
}
\newglossaryentry{active.portfolio}
{
  name={active portfolio},
  description={The active portfolio $A$ of portfolio $W$ against \gls{benchmark.portfolio} $B$ is the portfolio $A = W - B$}
}
\newglossaryentry{adjusted.nominal.gain}
{
  name={adjusted nominal gain},
  description={In terminology developed for this book, the \emph{adjusted nominal gain} is the total amount won (or if lost, a negative value) by a trading system when a method for determining trade size is used. See also \gls{nominal.gain}}
}
\newglossaryentry{ADR}
{
  name={ADR},
  description={An acronym for \emph{American Depository Receipts}. These are stocks of foreign companies that have listings on American stock exchanges and trade in dollars. Cash flows of these stocks are converted into dollars at the prevailing foreign exchange rate}
}
\newglossaryentry{affect}
{
  name={affect},
  description={Affect refers a specific quality of ``goodness'' or ``badness'' produced by \gls{System1} and in the words of Paul Slovic and colleagues, ``(i) experienced as a feeling state (with or without consciousness) and (ii) demarcating a positive or negative quality of a stimulus.'' Ordinary decision makers make heavy use of affect to make ``snap judgments'' based on feelings of like or dislike. Such feelings are conditioned by experience, and even a single episode is enough to condition an affect response}
}
\newglossaryentry{affect.heuristic}
{
  name={affect heuristic},
  description={A decision rule in which outcomes are parsed into the simple categories ``good' and ``bad'' using \gls{System1} heuristics. especially emotions (e.g., liking, hating), or moods (e.g. happy, sad, mad). In the common vernacular, ``gut feelings'' is the closest correspondent}
}
\newglossaryentry{aggregation}
{
  name={aggregation},
  description={As used in the \glslink{efficient.market.hypothesis}{EMH}, refers to the process of information dissemination when some players are more informed than others, i.e. to the manner in which prices adjust to reflect the information of a minority, the \glspl{.insider}. Also called \gls{information.aggregation}}
}
\newglossaryentry{agreement}
{
  name={agreement},
  description={In the \glspl{efficient.market.hypothesis}, the requirement that investors must agree on the correct prices of assets}
}
\newglossaryentry{algorithmic.strategy}
{
  name={algorithmic strategy},
  description={A \gls{strategy} that a part of a \gls{mechanical.trading.system}}
}
\newglossaryentry{algorithmic.trade}
{
  name={algorithmic trade},
  description={(1) In a \gls{PPGS.renewal.process}, all the \glspl{transaction} occuring from the time of the \gls{founding.transaction} until that transaction is closed, and (2) the part of a \gls{mechanical.trading.system}, usually computer-assisted, that performs computations to buy or sell}
}
\newglossaryentry{algorithmize}
{
  name={algorithmize},
  description={The process of taking a verbal or symbolic ``idea'' and converting it to an algorithm. For example: ``sort a set of numbers'' is algorithmized by a bubble sort}
}
\newglossaryentry{ambiguity.aversion}
{
  name={ambiguity aversion},
  description={\emph{Ambiguity aversion} is a preference for known risks over unknown risks, e.g. the Ellsberg paradox. Also known as \emph{uncertainty aversion}}
}
\newglossaryentry{anchor}
{
  name={anchor},
  description={A decision heuristic that uses some piece of information, even irrelevant information, as a reference point for a decision}
}
\newglossaryentry{anchoring.and.adjustment}
{
  name={anchoring \& adjustment},
  description={An effect in which the \gls{associative.machine} uses ambient information, including irrelevant information, to make a decision}
}
\newglossaryentry{anomaly}
{
  name={anomaly},
  description={In the context of the \gls{efficient.market.hypothesis}, a market phenomenon that has no apparent \acrshort{aEMH} explanation, i.e., an apparent contradition to the \acrshort{aEMH}},
  plural={anomalies}
}
\newglossaryentry{arb}
{
  name={arb},
  description={Nickname for an \gls{arbitrageur}}
}
\newglossaryentry{arbbed.out}
{
  name={arbbed out},
  description={Referring to a mispricing that has been corrected through \gls{arbitrage}}
}
\newglossaryentry{arbbing}
{
  name={arbbing},
  description={The process of engaging in \gls{arbitrage}}
}
\newglossaryentry{arbitrage}
{
  name={arbitrage},
  description={A strategy that buys and sells the same or statistically similar securities at advantageously different prices, e.g. index \gls{arbitrage}, \glslink{pairs.trading.strategy}{pairs trading}, options delta-neutral hedging. See also \gls{pure.arbitrage} and \gls{statistical.arbitrage}}
}
\newglossaryentry{Arbitrage.Pricing.Theory}
{
  name={Arbitrage Pricing Theory},
  description={A theory that extends the \glslink{Capital.Asset.Pricing.Model}{CAPM} by assuming that several factors besides the market are needed to model asset returns}
}
\newglossaryentry{arbitrageur}
{
  name={arbitrageur},
  description={A trader whose dominant strategies are \glspl{arbitrage}}
}
\newglossaryentry{ariely}
{
  name={Dan Ariely},
  description={Well-known researcher in behavioral finance, author of ``Predictably Irrational''}
}
\newglossaryentry{artificial.mathematical.game}
{
  name={artificial mathematical game},
  description={A stuctured contest between several mathematical entities (players) who engage strategically to determine their payoffs. See also \gls{game.theory}}
}
\newglossaryentry{asexual}
{
  name={asexual},
  description={An asexual organism does not require two parents to reproduce, but clones itself instead}
}
\newglossaryentry{aspiration.level}
{
  name={aspiration level},
  description={In the \gls{spa} theory, a means of assessing the attractiveness of a lottery using the probability that it is not below an aspiration level $\alpha$: $A = P[ V \ge \alpha]$, where $V$ equals the security-potential value}
}
\newglossaryentry{associative.machine}
{
  name={associative machine},
  description={A term used in this book in reference to the \gls{System1} method of storing memories in a network where they are connected with other memories that occurred at about the same time, at about the same place, or with respect to other salient characteristics}
}
\newglossaryentry{asymptotic.growth.rate}
{
  name={asymptotic growth rate},
  description={A \gls{geometric.growth.rate} which a time series or stochastic process approaches almost surely}
}
\newglossaryentry{attribute.substitution}
{
  name={attribute substitution},
  description={In the making of intuitive judgments, the operation of substituting one attribute for another in order to facilitate a decision. For example, instead of answering ``Which stock has the greatest prospects?", the question could morph into ``Which stock to I like the most?" See also \gls{availability} and the \gls{affect.heuristic}}
}
\newglossaryentry{Axiom.Of.Archimedes}
{
  name={Axiom of Archimedes},
  description={An axiom of utility theory (also called the \emph{\gls{Continuity.Axiom}}) which asserts that if $B$ is an event that is ``sandwiched'' between $A$ and $B$ ($A \succ B$ and $B \succ C$), then there exists a $p \in (0,1)$ such that $B$ and the prospect $(p, A; (1-p), C)$ are equivalent}
}
\newglossaryentry{availability}
{
  name={availability},
  description={In the psychology of decision making, referring to the (possibly biased) information used as input to decision making}
}
\newglossaryentry{availability.heuristic}
{
  name={availability heuristic},
  description={A decision heuristic that chooses based on readily available information}
}

\newglossaryentry{back.end}
{
  name={back end},
  description={A term used for the longer maturity instruments of the yield curve, typically ones with a maturity of $5$ years or more}
}
\newglossaryentry{backtesting}
{
  name={backtesting},
  description={The process of applying quantative methods to a trading system's historical or simulated performance to determine if it is worth trading}
}
\newglossaryentry{BnL.purely.idiosyncratic.model}
{
  name={B\&L purely idiosyncratic model},
  description={A \acrlong{abnl} evolutionary model in which each parent in each generation has a random number of offspring independently of all others}
}
\newglossaryentry{BnL.systematic.model}
{
  name={B\&L systematic model},
  description={A \acrlong{abnl} evolutionary model in which each parent has the same number of offspring as all others of that generation}
}
\newglossaryentry{bankroll}
{
  name={bankroll},
  description={Literally, in the vulgar vernacular of gamblers and grifters, a bulging roll of paper money used to finance wagers. Polite company drops the reference to paper money, preferring the euphemism \emph{investment capital}}
}
\newglossaryentry{bargaining.problem}
{
  name={bargaining problem},
  description={A game theory problem in which two players negotiate a fair price when there exists an interval of prices favorable to each. A solution was given by John Nash in $1950$}
}
\newglossaryentry{basis.point}
{
  name={basis point},
  description={In fixed income parlance with respect to interest rates, one one-hundredth of a percent, $0.01\% = 0.0001$}
}
\newglossaryentry{basket.of.stocks}
{
  name={basket of stocks},
  description={A portfolio of stocks that typically tracks a benchmark stock index, and which can be sold as part of a single trade}
}
\newglossaryentry{basket.trade}
{
  name={basket trade},
  description={A \gls{trade} of a \gls{basket.of.stocks}. Market makers specializing in trading entire baskets are available to instutional traders. Exghange traded funds \acrshort{aETF}'s are similar to \emph{basket trades} except that the ownership of the underlying securities is held in trust, not by the owner of \emph{ETF} shares}
}
\newglossaryentry{Bayes'rule}
{
  name={Bayes' rule},
  description={A property of probabilities that for events $A$ and $B$, relates $P[B|A]$ to $P[A]$, $P[B]$ and $P[A|B]$ by the formula, 
\[
  P[B|A] = P[A|B] P[B] / P[A]
\]},
}
\newglossaryentry{behavioral.economics}
{
  name={behavioral economics},
  description={The study of economic phenomena that arise from the joint behavior of human economic actors as opposed to rational automatons. BF has roots in anthropology, marketing, psychology, social psychology and sociology}
}
\newglossaryentry{behavioral.finance}
{
  name={behavioral finance},
  description={The study of finance which examines how market phenomena arise from the interactions of its human participants. BF has untrustworthy roots in anthropology, psychology, social psychology and sociology}
}
\newglossaryentry{belief.revision.bias}
{
  name={belief revision bias},
  description={The tendency of people to revise prior beliefs slower than would be optimal by \gls{Bayes'rule}. Also known as the \gls{conservatism.bias}}
}
\newglossaryentry{benchmark}
{
  name={benchmark},
  description={In finance, a numerical standard against which the performance an investment vehicle is measured, e.g., the S\&P 500 Industrial Index, the Russell 2000 Index or LIBOR. In applications, benchmarks are usually diversified indices created by financial exchanges, investment houses or financial information providers}
}
\newglossaryentry{benchmark.portfolio}
{
  name={benchmark portfolio},
  description={In finance, a portfolio of tradable securities that serves a benchmark} 
}
\newglossaryentry{bias}
{
  name={bias},
  description={In the theory of statistical estimation, an estimator is \emph{unbiased} if its expected value equals the parameter being estimated, \emph{biased} otherwise}
}
\newglossaryentry{binary.channel}
{
  name={binary channel},
  description={In \glslink{shannon}{Shannon's} theory of communication, a system that has an input alphabet with two symbols, and output alphabet with two symbols, a ``channel'' that sends input to output symbols, and conditional probabilities that each output symbol will be received given that an input symbol is sent}
}
\newglossaryentry{binary.relation}
{
  name={binary relation},
  description={In mathematics, a binary relation on a set $\Omega$ is a set of ordered pairs $(x,y)$, where $x,y \in \Omega$}
}
\newglossaryentry{bouchaud}
{
  name={Jean-Phillipe Bouchaud},
  description={A French physicist born in 1962. He is founder and Chairman of Capital Fund Management (CFM) and professor of physics at \'{E}cole polytechnique}
}
\newglossaryentry{boundary.bias}
{
  name={boundary bias},
  description={A general problem with methods that perform smoothing of data in a bounded region. Toward the boundaries (edges) of the region, the dearth of data in a symmetric region around the prediction point reduces accuracy. In time series, for example, the ``present'' is a boundary, and the bias occurs because the future values are not available to regularize the smooth},
  plural={boundary biases}
}
\newglossaryentry{bounded.rationality}
{
  name={bounded rationality},
  description={A viewpoint that humans mostly behave rationally, but are constrained by limited capacities for mental computing and by difficulties in quantifying uncertainty}
}
\newglossaryentry{Buffett.Warren}
{
  name={Warren Buffett},
  description={The must famous investor of the latter part of the 20^{th} century and early 21^{st} century}
}

\newglossaryentry{calendar.time}
{
  name={calendar time},
  description={Referring to time measured in calendar units, e.g., July 2, 2014 at 11:06:42}
}
\newglossaryentry{cap.and.trade}
{
  name={cap-and-trade},
  description={A system of allocating pollution rights by issuing tradable licenses to pollute}
}
\newglossaryentry{capacity}
{
  name={capacity},
  description={Let $X$ be a set and $\mathscr{E}(X)$ be the set of subsets of $X$ (its power set). A capacity $W$ on $X$ is a function that maps $\mathscr{E}(X)$ to $[0,1]$ and has properties  $W(\phi) = 0$, $W(X) = 1$ and $W(A) \le W(B)$ if $A \subset B$}
}
\newglossaryentry{Capital.Asset.Pricing.Model}
{
  name={Capital Asset Pricing Model},
  description={An efficient markets model for pricing securities}
}
\newglossaryentry{capital.gains.overhang}
{
  name={capital gains overhang},
  description={For a tradable asset, the aggregate gains or losses, i.e., the sum over all (unrealized) differences between the current price and investors' purchase prices}
}
\newglossaryentry{CAR}
{
  name={CAR},
  description={Cumulative Abnormal Return}
}
\newglossaryentry{carry}
{
  name={carry},
  description={The \emph{carry} of an asset is the return obtained from holding it. Carry is positive if owning the asset earns money and negative if it costs money. For example, owning a note that pays $3\%$ on principal has a carry of $3\%$ minus the usually negligible costs (i.e. service charges on a brokerage account) of maintaining the position}
}
\newglossaryentry{carry.trade}
{
  name={carry trade},
  description={A strategy that finances an pspeculative by trades borrowing in a currency that has relatively low interest rates}
}
\newglossaryentry{cardinal.utility}
{
  name={cardinal utility},
  description={See \gls{utility.theory}}
}
\newglossaryentry{certainty.equivalent}
{
  name={certainty equivalent},
  description={In utility theory involving monetary prizes, the minimum cash amount that is acceptable to forego the lottery}
}
\newglossaryentry{cheater}
{
  name={cheater},
  description={A person or trading entity that engages in illegal activity. ``Unethical'' activity is excluded since in financial theory, such activity does not exist}
}
\newglossaryentry{choice.architecture}
{
  name={choice architecture},
  description={A concept in applied \gls{behavioral.finance} that highlights the impact of the framing of choices on the outcomes. Choice architecture is an application of \glslink{prospect.theory}{prospect theory's} demonstration that framing of a \gls{decision.problem} can decisively affect the outcome}
}
\newglossaryentry{closed-end.discount.puzzle}
{
  name={Closed End Discount Puzzle},
  description={An efficient market anomaly that has no good rational explanation. Closed end funds trade at steep discounts to book value most of the time, have increased issuance during bull markets and converge to book upon close-ending}
}
\newglossaryentry{closing.transaction}
{
  name={closing transaction},
  description={In a \gls{PPGS.renewal.process}, a \gls{transaction} that exits an \gls{algorithmic.trade} resulting in no residual position}
}
\newglossaryentry{cluster.analysis}
{
  name={cluster analysis},
  description={An analysis that uses a data-analytic methods for identifying groups within a multidimensional dataset}
}
\newglossaryentry{cognitive.dissonance}
{
  name={cognitive dissonance},
  description={The mental stress caused by either holding, or being confronted by, a belief that is incoherent with those of the \gls{associative.machine}}
}
\newglossaryentry{coefficient.of.absolute.risk.aversion}
{
  name={coefficient of absolute risk aversion},
  description={For \gls{utility.function} $U$, the measure $r(x) = -U''(x) / U'x)$}
}
\newglossaryentry{coefficient.of.relative.risk.aversion}
{
  name={coefficient of relative risk aversion},
  description={For \gls{utility.function} $U$, the measure $r(x) = -x U''(x) / U'x)$}
}
\newglossaryentry{cognition}
{
  name={cognition},
  description={In psychology, referring to individuals' mental processes that process information, which includes, inter alia those of attention, memory, comprehending, learning and using language, calculating, reasoning, problem solving, and decision making}
}
\newglossaryentry{coherent}
{
  name={coherent},
  description={In \gls{System1} perception, pattern recognition intended to match most closely the existing mental constructs of the \gls{associative.machine}. Coherence is fundamental to \glslink{System1}{System 1's} decision processes}
}
\newglossaryentry{competition.for.the.second.move}
{
  name={competition for the second move},
  description={Since there are often benefits to moving second in two-player games, due to the information revealed in the first mover's choice, there can be a ``competition'' to move second}
}
\newglossaryentry{Complete.Ordering}
{
  name={Complete Ordering},
  description={An axiom of utility theory that requires that for any two events $A$ and $B$, at least one of $A \succeq B$ and $B \succeq A$ must be true}
}
\newglossaryentry{complexity}
{
  name={complexity},
  description={The quality or state of a system that is not simple; in this work, of a dynamic system whose laws of motion are too complicated to be quantified using known laws}
}
\newglossaryentry{component.decomposition}
{
  name={component decomposition},
  description={A functional decomposition of multidimensional data into simpler, usually $1$-dimensional factors. Principal components analysis is an example}
}
\newglossaryentry{Compound.Equivalence}
{
  name={Compound Equivalence},
  description={An axiom of utility theory that requires that lotteries of lotteries must follow the multiplication law of probabilities}
}
\newglossaryentry{concave.strategy}
{
  name={concave strategy},
  description={A \gls{strategy} that sells when price is going up and buys when price is going down},
  plural={concave strategies}
}
\newglossaryentry{confirmation.bias}
{
  name={confirmation bias},
  description={The tendency to process new information in a way that confirms one's point of view. The bias arises from the \glslink{associative.machine}{associative machine's} preference for \gls{coherent} explanations}
}
\newglossaryentry{conservatism.bias}
{
  name={conservatism bias},
  description={The tendency of people to revise prior beliefs slower than would be optimal by \gls{Bayes'rule}. Also known as the \gls{belief.revision.bias}}
}
\newglossaryentry{Continuity.Axiom}
{
  name={Continuity Axiom},
  description={An axiom of utility theory (also called the \emph{Archimedian axiom}) which asserts that if $B$ is an event that is ``sandwiched'' between $A$ and $B$ ($A \succ B$ and $B \succ C$), then there exists a $p \in (0,1)$ such that $B$ and the prospect $(p, A; (1-p), B)$ are equivalent}
}
\newglossaryentry{continuous.auction.market}
{
  name={continuous auction market},
  description={A market in which participants may place orders to buy, sell or cancel at any time during the venue's hours of operation. Transactions in a continuous auction market occur only when a buyer meets a seller's price or seller meets a buyer's price}
}
\newglossaryentry{contrarian}
{
  name={contrarian},
  description={An investor who follows contrarian strategies. Contrarian strategies posit that financial instruments go through periods of being \gls{overbought} and \gls{oversold} and that these can be detected, yielding trading \glspl{edge}}
}
\newglossaryentry{convex.strategy}
{
  name={convex strategy},
  description={A \gls{strategy} that buys when price is going up and sells when price is going down},
  plural={convex strategies}
}
\newglossaryentry{coordination.game}
{
  name={coordination game},
  description={A game in which individuals must cooperate to achieve a jointly favorable outcome, but in any agreement to cooperate, most achieve less than their maximal outcome}
}
\newglossaryentry{cost.of.carry}
{
  name={cost of carry},
  description={The negative of \gls{carry}. For example, selling a note that pays $3\%$ on principal has a cost of carry of $3\%$ plus the usually negligible costs (i.e. service charges on a brokerage account) of maintaining the position}
}
\newglossaryentry{counter.cumulative.distribution.function}
{
  name={counter-cumulative distribution function},
  description={The function $S(x) = F_{\!_{>}}(x) = 1 - F_{\!_{\le}}(x)$, where $F_{\!_{\le}}(x)$ is a \gls{cumulative.distribution.function}. Also called a \gls{survival.function}}
}
\newglossaryentry{crowding}
{
  name={crowding},
  description={Crowding is a phenomenon in which ``too many'' \glspl{arb} attempt to exploit a mispricing, resulting in a ``tragedy of the commons'' effect. In some cases, the result is a market reversal in which the \gls{arbitrage} moves violently against the  \glspl{arb}, inflicting upon them significant losses. Some cases in which this was thought to have happened: (1) the collapse of the hedge fund \gls{LTCM} and (2) the Hedge Fund Crisis of August, 2007}
}
\newglossaryentry{cumulative.distribution.function}
{
  name={cumulative distribution function},
  description={The function $F_{\!_{\le}}(x) = P[X \le x]$. Often called simply the \gls{distribution.function}}
}
\newglossaryentry{cumulative.prospect.theory}
{
  name={cumulative prospect theory},
  description={Cumulative prospect theory was introduced by Kahneman and Tversky in 1992 to address certain problems with the original \gls{prospect.theory}, most notably its failure to preserve stochastic dominance}
}
\newglossaryentry{curse.of.dimensionality}
{
  name={curse of dimensionality},
  description={A term that refers to various problems that arise when analyzing data from high-dimensional spaces. For example, when variables representing dimensions are stochastically independent, then almost all the probability within a ball of radius $R$ will lie in a thin shell between $R$ and $R - \epsilon$}
}

\newglossaryentry{data.mining}
{
  name={data mining},
  description={In data science, techniques that attempt to find patterns in (generally large) datasets. Such techniques are succeptible to the \gls{problem.of.multiplicity}}
}
\newglossaryentry{decision.science}
{
  name={decision science},
  description={The study of human decision making in which decisions of other parties can be ignored, often called ``games against nature.'' Three branches are often distinguished: (1) normative, the ``correct'' way to make decisions, (2) descriptive, the way humans decide in reality, and (3) prescriptive, the study of improving decision making}
}
\newglossaryentry{decision.problem}
{
  name={decision problem},
  description={A mathematical problem that abstracts decision making for a player when no strategic behavior by other players is possible}
}
\newglossaryentry{decision.theory}
{
  name={decision theory},
  description={A term for normative \gls{decision.science}}
}
\newglossaryentry{decoding}
{
  name={decoding},
  description={The process of recovering an encoded message}
}
\newglossaryentry{decrementing.transaction}
{
  name={decrementing transaction},
  description={In a \gls{PPGS.renewal.process}, a \gls{transaction} for which the number of shares, units or contracts of the prior \gls{position} is decreased}
}
\newglossaryentry{decumulative.distribution.function}
{
  name={decumulative distribution function},
  description={The function $D(x) = P[ X \ge x]$. It is related to the \gls{counter.cumulative.distribution.function} $S(x)$ by $D(x) = S(x) + P[X = x]$ The decumulative distribution function differs from the \gls{counter.cumulative.distribution.function} only if there are probability mass points, e.g. discrete distributions}
}
\newglossaryentry{defined.benefit.plan}
{
  name={defined-benefit plan},
  description={An annuity that promises payments calculated by a specific formula, usually based on years of service and income earned during that time. Persons in a defined-benefit plan have little choice about how to invest; their main decision is when to begin withdrawing benefits}
}
\newglossaryentry{defined.contribution.plan}
{
  name={defined-contribution plan},
  description={An completely portable savings annuity that requires an employee to elect how much to contribute, what investments to select and when to begin withdrawing}
}
\newglossaryentry{degenerate.random.variable}
{
  name={degenerate random variable},
  description={A \gls{random.variable} that assumes a single value with probability $1$}
}
\newglossaryentry{degree.of.diffusion}
{
  name={degree of diffusion},
  description={An informal term that describes the extent to which a strategy is known by the community of traders. A strategy with low degree of diffusion has few followers, but the \gls{strategy.capacity} must be sufficient to present \gls{arbitrage} opportunities}
}
\newglossaryentry{descriptive.decision.science}
{
  name={descriptive decision science},
  description={See \gls{decision.science}}
}
\newglossaryentry{dimensional.reduction}
{
  name={dimensional reduction},
  description={The reduction of the complexity of a dataset or problem that has a large number of variables by ideally identifying a small number of functions of those variables that explain salient features}
}
\newglossaryentry{direction}
{
  name={direction},
  description={In mathematics, a vector having unit length}
}
\newglossaryentry{discretionary.trade}
{
  name={discretionary trade},
  description={A \gls{trade} initiated by a human trader using judgment, not a purely mechanical algorithm}
}
\newglossaryentry{disposition.effect}
{
  name={disposition effect},
  description={The tendency of investors to sell winners to early and hold losers too long}
}
\newglossaryentry{distortion.factor}
{
  name={distortion factor},
  description={Any exogenous news, constraints on trading, widely held beliefs, or widely used strategies that effect \glspl{price.impact}. For example, the fact that mutual funds cannot short stocks suggests that prices might be higher than they would be if shorting were allowed. Same as \gls{price.distorter}}
}
\newglossaryentry{distribution.function}
{
  name={distribution function},
  description={The function $F_{\!_{\le}}(x) = P[X \le x]$. Also called the \gls{cumulative.distribution.function}}
}
\newglossaryentry{Dot-com.Bubble}
{
  name={Dot-com Bubble},
  description={A remarkable modern bubble ($1996$-$2000$) powered by the fantasy of a ``new economy.'' The story line was that ``.com'' (read: ``dot com'') companies would in time eradicate conventional retail storefronts by offering products online. In a scant four years, this craze caused the Nasdaq composite index to quadruple, going from about $1,200$ to over $5,000$. But remarkably, an abrupt crash did not occur; instead, there occurred a slow, steady and punishing decline back to about $1,500$}
}
\newglossaryentry{dominant.paradigm}
{
  name={dominant paradigm},
  description={A \gls{market.paradigm} which is the best expression of market sentiment},
}
\newglossaryentry{dominated}
{
  name={dominate},
  description={In game theory, a strategy for a player is said to be \emph{dominated} is there is another strategy that always produces better payoffs. One solution method in games consists of successive elimination of dominated strategies, and chosing one of the remaining ones. Thie  method has limited usefulness for solving games because it seldom produces a unique solution}
}
\newglossaryentry{downsizing.transaction}
{
  name={downsizing transaction},
  description={In a \gls{PPGS.renewal.process}, a \gls{transaction} that decreases the absolute value of a current, nonzero \gls{position} without reducing it to zero}
}
\newglossaryentry{drawdown}
{
  name={drawdown},
  description={If not explicitly stated otherwise, a peak-to-trough drawdown of a financial time series is its greatest loss from a previous high usually expressed as a percentage. For example, a series that had a previous high of $1000$ and a lowest low later of $800$ would have at least a $20\%$ peak-to-trough drawdown}
}
\newglossaryentry{drawdown.loss.function}
{
  name={drawdown loss function},
  description={For a real-valued time series $\{X_s\}_{s=1}^{s=t}$ of gains, the largest cumulative loss $\mathcal{L}(X,t)$ from a previous high, i.e. for $V_s = \sum_{u=1}^{u=s} X_u$, $ s \ge 1$ and $V_0 = 0$,
\[
  \mathcal{L}(X,t) = \underset{0 \le s \le t}{max} \left\{ (\underset{0 \le u \le s}{max} V_u) \, - \, V_s ) \right\}.
\]
Note that $\mathcal{L}(X,t) \ge 0$}
}
\newglossaryentry{driftless.random.walk}
{
  name={driftless random walk},
  description={A random walk $X_t = \mu + X_{t-1} + U_t$ for which $\mu = 0$}
}
\newglossaryentry{DSSW.Model}
{
  name={DSSW Model},
  description={A model of bubbles in which (1) \glspl{fundamentalist} correct a mispricing, creating the illusion of a trend, (2) \glspl{rational.speculator} push price away from fundamental value after fundamental value is restored, and continue until they can offload it to \glspl{noise.trader}. In this work, called \gls{legal.pump.and.dump}}
}
\newglossaryentry{durability.bias}
{
  name={durability bias},
  description={Refers to the tendency for people to overestimate the persistence of a positive or negative event, episode or circumstance}
}

\newglossaryentry{ecology}
{
  name={ecology},
  description={In finance, an ecology at a moment in time includes both the \gls{strategic.ecology} and market structure, in the sense of a game that has rules, venues and technology within which strategies can be enacted. Also called \gls{market.ecology}}
}
\newglossaryentry{econ}
{
  name={Econ},
  description={A \acrshort{avnm} financial decision maker. Also called \gls{homo.economicus} (\acrshort{aHE})}
}
\newglossaryentry{econophysics}
{
  name={econophysics},
  description={The study of economic phenomena by persons trained in the sciences, whether physicists or not, using techniques borrowed from those sciences}
}
\newglossaryentry{edge}
{
  name={edge},
  description={A gambling term for the expected gain or loss from a bet. A bet with positve expectation has ``edge'', and one that is not positive, has no ``edge''}
}
\newglossaryentry{efficient.market.hypothesis}
{
  name={Efficient Market Hypothesis},
  description={The economic theory that markets incorporate all available information making it impossible for investors to consistently earn excess profits}
}
\newglossaryentry{EMH.1}
{
  name={EMH.1},
  description={Rationality: Investors are mostly rational}
}
\newglossaryentry{EMH.2}
{
  name={EMH.2},
  description={Cancellation: To the extent that investors are not rational, they make random trading decisions that in aggregate produce random deviations from fair prices}
}
\newglossaryentry{EMH.3}
{
  name={EMH.3},
  description={Arbitrage: To the extent that the random deviations of \gls{EMH.2} are sufficiently large, rational traders enter the market and correct them}
}
\newglossaryentry{EMH-S}
{
  name={EMH-S},
  description={Strong form of the \gls{efficient.market.hypothesis}: It is impossible to earn superior \glspl{risk-adjusted.return} given public (i\gls{EMH-W} and \gls{EMH-SS}) and private (\emph{\gls{.insider}}) information}
}
\newglossaryentry{EMH-SS}
{
  name={EMH-SS},
  description={Semi-strong form of the \gls{efficient.market.hypothesis}: It is impossible to earn superior \glspl{risk-adjusted.return} given publicly available information, which includes \gls{EMH-W} information plus data such as \gls{.insider} trader SEC reports of the purchases and sales of their stocks by corporate officers, earnings and dividends surprises, and accounting and auditing reports}
}
\newglossaryentry{EMH-W}
{
  name={EMH-W},
  description={Weak form of the \gls{efficient.market.hypothesis}: It is impossible to earn superior \glspl{risk-adjusted.return} given knowledge of past prices, returns, volume and open interest alone}
}
\newglossaryentry{empirical.distribution.function}
{
  name={empirical distribution function},
  description={For real numbers $x_1, x_2, \ldots, x_n$, the empirical distribution function $\hat{F}(x)$ is defined by
  \[
    \hat{F}(x) = \frac{1}{n} \sum_{i=1}^{i=n} I( x \ge x_i ) = \frac{1}{n} \{ \# i \, | \, x_i \le x \},
  \]
  where $I(A)$ is the indicator function of $A$. When $X_1, X_2, \ldots, X_n$ is a random sample from \gls{cumulative.distribution.function} $F(x)$, $\hat{F}(x)$ converves uniformly to $F(x)$ (Glivenko-Cantelli Theorem)}
}
\newglossaryentry{encoding}
{
  name={encoding},
  description={The transformation of a message into a code}
}
\newglossaryentry{endogenous.event}
{
  name={endogenous event},
  description={In systems theory, an event that can be explained by the system; it occurs ``inside'' it, or rather, as a consequence of it. Opposed to an \gls{exogenous.event}}
}
\newglossaryentry{endogenous.market.distortion}
{
  name={endogenous market distortion},
  description={A predictable behavior of markets due to the interaction of strategies. Example: bubbles, crashes and \gls{crowding}}
}
\newglossaryentry{endowment.effect}
{
  name={endowment effect},
  description={A bias in which a person values the same object more if it is owned than if it is not}
}
\newglossaryentry{ensemble}
{
  name={ensemble},
  description={The set of all paths (histories) that a stochastic process can follow}
}
\newglossaryentry{equivalence}
{
  name={equivalence},
  description={In \gls{utility.theory}, equivalence of events $A$ and $B$ is expressed as the requirement that $A \succeq B$ and $B \succeq A$. Such events are also said to be \emph{equivalent}. See also \gls{indifference}}
}
\newglossaryentry{ergodic}
{
  name={ergodic},
  description={A stochastic process whose ensemble averages are almost surely the same as its asymptotic time averages}
}
\newglossaryentry{event-precipitated.trend}
{
  name={event-precipitated trend},
  description={A trend that occurs due to an event}
}
\newglossaryentry{event.time}
{
  name={event time},
  description={Referring to time measured by the occurrence of events}
}
\newglossaryentry{exchange.traded.fund}
{
  name={exchange traded fund},
  description={A type of mutual fund in which shares are claims on a published portfolio of stocks. These stocks are held in trust, and the value of shares determined in the usual way from the values of those stocks}
}
\newglossaryentry{exchange}
{
  name={exchange},
  description={A marketplace in which financial instruments such as equities, bonds, commodities, options and futures are traded; securities that trade on exchanges, which by law are required to report trading prices, are called \emph{listed}. In the U.S., the New York Stock Exchange (\acrshort{aNYSE}), American Stock Exchange (\acrshort{aAMEX}), and National Association of Securities Dealers Automatic Quotation System (\acrshort{anasdaq}) are the most established stock exchanges, but many new electronic exchanges have arisen since the $1990$'s. There is also an informal market that deals in over-the-counter securities}
}
\newglossaryentry{exogenous.event}
{
  name={exogenous event},
  description={In systems theory, an event that unexplained by the system; it occurs ``outside'' it. An earthquake that has adverse economic consequences, is an example of an exogenous event for financial markets. Opposed to an \gls{endogenous.event}}
}
\newglossaryentry{exogenous.market.distortion}
{
  name={exogenous market distortion},
  description={A \emph{exogenous market distortion} is a potentially predictable market behavior that results from constraint(s) due to  beliefs, strategies or institutional arrangements}
}
\newglossaryentry{exogenous.shock}
{
  name={exogenous shock},
  description={In financial theory, an event that occurs ``outside'' the market, and which therefore, is at least partly unanticipatible. Bad weather, eqrthquakes and tsunamis that have significant economic consequences are examples}
}
\newglossaryentry{exponential.distribution}
{
  name={exponential distribution},
  description={The exponential distribution having density $\lambda e^{- \lambda x}$ for $x > 0$}
}
\newglossaryentry{expected.utility}
{
  name={expected utility},
  description={In this work, the same as \gls{vNM.expected.utility.theory}}
}
\newglossaryentry{experimental.economics}
{
  name={experimental economics},
  description={A field developed by 2002 Nobel Laureate in Economics, Vernon Smith. Experimental econometricians conduct laboratory experiments to test the predictions of economic theories}
}
\newglossaryentry{experimental.data}
{
  name={experimental data},
  description={Data acquired in a controlled scientific experiment}
}
\newglossaryentry{expiration.pin.risk}
{
  name={expiration pin risk},
  description={The risk associated with an option conversion or reversal position when the underlying expiration price is its strike}
}
\newglossaryentry{exponential.moving.average}
{
  name={exponential.moving average},
  description={An exponential moving average with parameter $\beta$, $0 < \beta \le 1$ is a \gls{moving.average} having $\alpha_s = \beta (1 - \beta)^^s$, where $EMA_t = \sum_{s=0}^{s=\infty} \alpha_s X_{t-s}$}
}
\newglossaryentry{extreme.percentile.price.trending}
{
  name={extreme percentile price trending},
  description={A pricing anomaly in which stocks in extreme percentiles of past returns either continue or reverse those trends. An example of the former is stock \gls{momentum}, of the latter, the De Bondt-Thaler 3-5 year trend reversal in stocks}
}

\newglossaryentry{factor.model}
{
  name={factor model},
  description={A multiple regression or factor analysis model that identifies components, usually linear, that reduce the dataset dimensionality}
}
\newglossaryentry{falsifiability}
{
  name={falsifiability},
  description={As used with respect to a scientific hypothesis, the requirement that there exist experimental methods to show that with high probability, it is untrue}
}
\newglossaryentry{farmer}
{
  name={Doyne Farmer},
  description={Prominent physicist, econophysicist and entrepreneur formerly of the Santa Fe Institute, currently Co-Director, Complexity Economics, The Institute for New Economic Thinking at the Oxford Martin School}
}
\newglossaryentry{Fast.Frugal.Heuristics}
{
  name={Fast \& Frugal Heuristics},
  description={A term originated by \gls{gigerenzer} to describe domain-specific ``mental shortcuts'' for making decisions, e.g. the \emph{\gls{recognition.heuristic}} and the \emph{\gls{pick.the.best.heuristic}}}
}
\newglossaryentry{fertility}
{
  name={fertility},
  description={In population biology and demographics, the study of patterns of reproduction by members of a species}
}
\newglossaryentry{FIFO}
{
  name={FIFO},
  description={Abbreviation for \emph{First in, first out}. Used in queuing theory}
}
\newglossaryentry{founding.transaction}
{
  name={founding transaction},
  description={In a \gls{PPGS.renewal.process}, a \gls{transaction} that initiates a position, i.e. when the prior position is zero}
}
\newglossaryentry{float}
{
  name={float},
  description={For a stock, its float is the number of shares outstanding available without restrictions to the public}
}
\newglossaryentry{fractional.return}
{
  name={fractional return},
  description={In finance, given price $p$ at time $t'$ and $q$ at time $t'' > t'$, the fractional return is $(q-p)/p$}
}
\newglossaryentry{fragile}
{
  name={fragile},
  description={A condition in which a financial instrument or market has a non-trivial chance of a very large move. Not \glslink{robustness}{robust}}
}
\newglossaryentry{framing}
{
  name={framing},
  description={In \gls{prospect.theory}, a statement of the \gls{decision.problem} in terms of gains, losses, or neither. \acrshort{aKnT} showed framing to be critical in the decision process, and could even lead to preference reversals. For example, stating a problem in terms of gains could result in the opposite decisions from one stated in terms of losses}
}
\newglossaryentry{free.rider}
{
  name={free rider},
  description={In a public goods game, a player who contributes little or nothing but receives the benefits nonetheless}
}
\newglossaryentry{front.end}
{
  name={front end},
  description={A term used for the shorter maturity instruments of the yield curve, typically ones with a maturity of less than $5$ years}
}
\newglossaryentry{frontrunning}
{
  name={front running},
  description={The practice of brokers who place their orders ahead of large customer orders. The practice is illegal in many markets, a notable exception being foreign currency markets}
}
\newglossaryentry{full.disclosure}
{
  name={full disclosure},
  description={A non-standard term in \gls{game.theory} in which a player discloses that he or she will pursue a certain strategy with a probability of either $0$ or $1$}
}
\newglossaryentry{fundamental.analysis}
{
  name={fundamental analysis},
  description={The valuation of stocks based on financial statements, business analysis and industry comparisons},
  plural={fundamental analyses}
}
\newglossaryentry{fundamentalist}
{
  name={fundamentalist},
  description={In the \gls{DSSW.Model} of bubbles, a trader who buys and sells according to \gls{fundamental.analysis}. Many fundamentalists are precluded from shorting, e.g. mutual funds, while others are unwilling, since it is quite risky in real markets}
}
\newglossaryentry{Fundamental.Laws.of.Gambling}
{
  name={Fundamental Laws of Gambling},
  description={There are two Fundamental Laws of Gambling: (1) The Fundamental Law - Never bet unless you think you have an \gls{edge}, and (2) The Wagering Principle - Bet an amount appropriate for the edge}
}

\newglossaryentry{gambler's.fallacy}
{
  name={gambler's fallacy},
  description={The mistaken belief that if an event happens frequently in an \gls{i.i.d.} series, it is less likely to happen in the future} 
}
\newglossaryentry{gambler's.ruin}
{
  name={gambler's ruin},
  description={A gambler is said to be \emph{ruined} if his wagering capital is exhausted. In probability theory, \emph{gambler's ruin} problems involve one or more players with finite funds wagering repeatedly until one of them goes broke. In the simplest cases, gambler's ruin problems have closed form solutions (Feller, 1957)}
}
\newglossaryentry{gambling}
{
  name={gambling},
  description={The low enterprise of wagering on the outcomes of matters both small and large, and on events from the most trivial to the most consequential, with little regard for the outcomes excepting the gain or loss incurred. Not to be confused with \emph{investing}, which hides the enterprise behind a fa\c{c}ade of respectability}
}
\newglossaryentry{gambling.irrational}
{
  name={gambling irrational},
  description={Same as \gls{girrational}}
}
\newglossaryentry{gambling.rational}
{
  name={gambling rational},
  description={Same as \gls{grational}}
}
\newglossaryentry{gambling.system}
{
  name={gambling system},
  description={A gambling system for $\{X_t\}$ with respect to $\{Z_t\}$ is a family of real-valued functions $f_t(Z_t, Z_{t-1} , . . .)$ that specify how much to bet at time $t$ on the outcome at time $t + 1$ given the history $Z_t , Z_{t-1}, \ldots$}
}
\newglossaryentry{game.analysis}
{
  name={game analysis},
  description={The study of the best way to play rules-based, possibly stochastic, contests in which two or more participants (players) make decisions according to the rules, and receive payoffs as a function of a terminal outcome. Unlike \gls{game.theory}, game analysis is concerned with winning at games, not with theoretically perfect play}
}
\newglossaryentry{game.theory}
{
  name={game theory},
  description={The study of rules-based, possibly stochastic, contests in which two or more participants (players) make decisions according to the rules, and receive payoffs as a function of a terminal outcome. A \emph{solution} in game theory consists of strategies for the players that are in some sense optimal}
}
\newglossaryentry{gaussian.distribution}
{
  name={Gaussian distribution},
  description={The Gaussian (normal) distribution $N(\mu,\sigma^2)$ is a two-parameter distribution with $\mu$ the mean and $\sigma > 0$ the standard deviation. Its density is $\phi(x; \mu, \sigma) = (2 \pi \sigma^2)^{-1/2} exp(- ( (x - \mu)/ \sigma )^2)$, $x \in \mathbb{R}$},
  plural={Gaussian density}
}
\newglossaryentry{Gaussian.random.walk}
{
  name={Gaussian random walk},
  description={A random walk $X_t = \mu + X_{t-1} + U_t$ for which $U_t \sim N(\mu,\sigma^2)$}
}
\newglossaryentry{Gaussian.white.noise}
{
  name={Gaussian white noise},
  description={A \gls{white.noise} $\{X_t\}$ in which all $X_t$ have Gaussian distributions}
}
\newglossaryentry{geometric.growth.rate}
{
  name={geometric growth rate},
  description={For a positive time series $\{X_t\}$, $t = 0, 1, \ldots$, the almost sure limit $G$ defined by 
\[
  ln(X_n/X_0)/n \rightarrow G, \text{ as } n \rightarrow \infty, 
\]
provided it exists}
}
\newglossaryentry{Soros.George}
{
  name={George Soros},
  description={Arguably the most successful, but unquestionably the most famous hedge fund manager of the latter $20^{th}$ and early $21^{st}$ centuries}
}
\newglossaryentry{generalized.prisoner's.dilemma} 
{
  name={generalized prisoner's dilemma},
  description={A generalization of a 2 person \gls{prisoners.dilemma} to an $N$ player game in which a coalition $M$ faces a prisoner's dilemma against $N \ M$}
}
\newglossaryentry{gigerenzer}
{
  name={Gerd Gigerenzer},
  description={Psychologist who developed Fast \& Frugal Heuristics}
}
\newglossaryentry{girrational}
{
  name={girrational},
  description={Said of an investor who is not \gls{grational}}
}
\newglossaryentry{Global.Financial.Crisis}
{
  name={Global Financial Crisis},
  description={The financial conflagration that was clearly discernible by $2007$ when the subprime mortgage market melted down, by the fall of $2008$ becoming a worldwide crisis of all financial markets, finally ending in the spring of $2009$. The worst crisis since $1929$}
}
\newglossaryentry{granularity}
{
  name={granularity},
  description={Referring to the size of the time interval in regularly sampled data; \emph{coarse granularity} indicates sampling at long intervals, such as days, and \emph{fine granularity} indicates sampling at short intervals such as seconds}
}
\newglossaryentry{grational}
{
  name={grational},
  description={Said of an investor whose criterion for trades is that they have expected return of at least $r$ and probability of \glspl{drawdown} exceeding $d$ no more that $p$. See also \gls{girrational}}
}
\newglossaryentry{Great.Depression}
{
  name={Great Depression},
  description={The worldwide financial collapse that followed the stock market crash of $1929$, and lasted for nearly a decade},
}
\newglossaryentry{Grossman-Stiglitz.paradox}
{
  name={Grossman-Stiglitz paradox},
  description={An EMH paradox attributed to S. Grossman and J. Stiglitz which argues that costly research is (theoretically) unnecessary for an efficient market, but a market can't be made efficient without such research}
}

\newglossaryentry{Half.Cubic.Law}
{
  name={Half-Cubic Law},
  description={Laws discovered by the Stanley econophysicist group that relate aggregate stock impact, quantity traded and number of shares traded as power laws with exponents of either $3/2$ or $3$}
}
\newglossaryentry{halo.effect}
{
  name={halo effect},
  description={The tendency to view more favorably things associated with something or someone you like}
}
\newglossaryentry{heavy.tail}
{
  name={heavy tail},
  description={Referring to the large values of a positive probability distribution (the ``tail'') that has a slower than exponential die-off, i.e. $\underset{x \rightarrow \infty} {lim} \; e^{\lambda x} P[X >x] = \infty$. For distributions with both positive and negative values, a \emph{heavy upper tail} and \emph{heavy lower tail} are defined analogously}
}
\newglossaryentry{hedge.fund}
{
  name={hedge fund},
  description={A pooled investment vehicle whose legal structure allows the use of high  \gls{leverage} and shorting and which generally invests in liquid assets. Hedge funds differ from \glspl{mutual.fund}, which have limits on  \gls{leverage} and cannot short, and from private equity funds, which often invest in relatively illiquid assets}
}
\newglossaryentry{herding}
{
  name={herding},
  description={Imitative behavior by groups of animals or humans, where all or most individuals make the same decision at about the same time, move in the same direction at the same time, etc. During market crashes, for example, most people try to sell at the same time}
}
\newglossaryentry{high-frequency}
{
  name={high-frequency},
  description={Data having short, usually irregular, time spans between successive observations. In this work, it refers to data occurring, or sampled at, time periods less than or equal to $15$ minutes}
}
\newglossaryentry{high.kurtosis}
{
  name={high kurtosis},
  description={Said of a distribution that has a third central moment greater than $3$ (the kurtotis of a normal distribution)}
}
\newglossaryentry{hindsight.bias}
{
  name={hindsight bias},
  description={The tendency to view an event retrospectively as having greater probability than it was thought to have at the time. Also known as \gls{knew.it.all.along.effect}}
}
\newglossaryentry{holistic.perception}
{
  name={holistic perception},
  description={Perception that emphasizes a whole over its parts. As opposed to \gls{rule.based.perception}}
}
\newglossaryentry{homo.economicus}
{
  name={Homo Economicus},
  description={A \acrshort{avnm} financial decision maker. Also called \gls{econ}, abbreviated \acrshort{aHE}}
}
\newglossaryentry{homogeneity}
{
  name={homogeneity},
  description={In a time series, the property of having the same statistical behavior at all scales; self-similarity}
}
\newglossaryentry{homogeneous.random.walk}
{
  name={homogeneous random walk},
  description={A random walk $X_t = \mu + X_{t-1} + U_t$ with unrestricted $\mu$}
}
\newglossaryentry{hot.hand.fallacy}
{
  name={hot hand fallacy},
  description={The belief that someone who makes frequent correct choices will continue to do so in the future. This belief is obviously mistaken if the choices arise from an \gls{i.i.d.} series}
}
\newglossaryentry{hypothesis-forming.stage}
{
  name={hypothesis-forming stage},
  description={The stage of scientific inquiry in which hypotheses are proposed to explain experimental data}
}

\newglossaryentry{incrementing.transaction}
{
  name={incrementing transaction},
  description={In a \gls{PPGS.renewal.process}, a \gls{transaction} for which the number of shares, units or contracts of the prior \gls{position} is increased}
}
\newglossaryentry{identically.distributed}
{
  name={identically distributed},
  description={A set of random variables $\{X_i\}$ are 
               identically distributed if $P[X_i \le x] = 
               P[X_j \le x]$ for all $i \ne j$}
}
\newglossaryentry{illusory.correlation.causation}
{
  name={illusory correlation and causation},
  description={The commonly held (false) view that correlated events entail 
               causative relationships. It often results from a misunderstanding of regression analysis}
}
\newglossaryentry{Immediate.Adjustment}
{
  name={Immediate Adjustment},
  description={In the \glspl{efficient.market.hypothesis}, the requirement that prices adjust immediately and correctly to unanticipated news}
}
\newglossaryentry{independence}
{
  name={independence},
  description={A reference to a set of random variables that are \gls{independent}. Same as \gls{statistical.independence}}
}
\newglossaryentry{independent}
{
  name={independent},
  description={In its technical meaning from probability 
               theory, random variables $\{X_i\}$ are 
               (jointly) independent if the distribution 
               function of any finite subset $(X_{i_1}, 
               X_{i_2}, \ldots, X_{i_n})$ can be expressed
               as a product of marginal distributions:
               \[ 
                  P[X_1 \le x_1, X_2 \le x_2, \ldots, X_n \le x_n] = 
                  \prod_{i=1}^{i=n} P[X_i \le x_i]
               \]
               }
}
\newglossaryentry{independent.components.analysis}
{
  name={independent components analysis},
  description={A method of decomposing a multivariate 
               dataset into components that are 
               maximally independent. There are several 
               different methods of calculating such 
               components, depending on the measure of 
               independence used}
}
\newglossaryentry{independent.white.noise}
{
  name={independent white noise},
  description={A \gls{white.noise} $\{X_t\}$ in which
               for any choice of $t_1 < t_2 < \cdots < t_n$, 
               $X_{t_1}, X_{t_2}, \ldots, X_{t_n}$ are 
               jointly independent}
}
\newglossaryentry{information.aggregation}
{
  name={information aggregation},
  description={As used in the \glslink{efficient.market.hypothesis}{EMH}, refers to the process of information dissemination when some players are more informed than others, i.e. to the manner in which prices adjust to reflect the information of a minority, the \glspl{.insider}. Also called simply \gls{aggregation}}
}
\newglossaryentry{i.i.d.white.noise}
{
  name={i.i.d. white noise},
  description={A \gls{white.noise} $\{X_t\}$ for which 
               all $X_t$ have the same distribution and 
               for any choice of $t_1 < t_2 < \cdots < t_n$, 
               $X_{t_1}, X_{t_2}, \ldots, X_{t_n}$ are 
               jointly independent}
}
\newglossaryentry{i.i.d.}
{
  name={i.i.d.},
  description={An abbreviation meaning ``independent and 
               identically distributed,'' referring to a 
               random sample or time series  $\{X_t\}$ 
               for which each $X_t$ has the same probability 
               distribution and any subset $X_{t_1}, X_{t_2}, 
               \ldots, X_{t_k}$ of variables are 
               independent}
}
\newglossaryentry{indifference}
{
  name={indifference},
  description={In \gls{utility.theory}, indifference between events $A$ and $B$ is expressed as the requirement that $A \succeq B$ and $B \succeq A$. An individual is said to be \emph{indifferent} to such events. See also \gls{equivalence}}
}
\newglossaryentry{information.ratio}
{
  name={information ratio},
  description={The \emph{information ratio} for return series $R_t$,
               $t = 1, 2, \ldots, n$ is its mean divided by its 
               standard deviation. The information ratio is similar to the 
               \gls{Sharpe.ratio} but is more useful in the Grinold \& 
               Kahn evaluation of active management}
}
\newglossaryentry{information.set}
{
  name={information set},
  description={For a time series $\{X_t\}$ at time $t'$, 
               a set $I_{t'}$ that gives data known at $t'$.  
               Generally used to form a conditional 
               distribution $X_{t'+1} \, | \, I_{t'}$ for the 
               next-period value, $X_{t'+1}$}
}
\newglossaryentry{.insider}
{
  name={insider},
  description={An individual who has access to non-public information, generally through channels of position, power or wealth}
}
\newglossaryentry{insider.trading}
{
  name={insider trading},
  description={The deplorable practice of trading on non-public information in advance of its market moving impact. At this writing, insider trading is illegal in the United States, but not in Monaco}
}
\newglossaryentry{intermittency}
{
  name={intermittency},
  description={In the context of this book, intermittency refers to the irregular onset of price volatility, in which there are alternating periods of high and low volatility.}
}
\newglossaryentry{intrinsic.value}
{
  name={intrinsic value},
  description={The value of a company or assed based on a balanced appraisal of all its characteristics and aspects, both tangible and intangible. \Gls{fundamental.analysis} strategies base their trading on estimated intrinsic values compared to market values}
}
\newglossaryentry{IPO}
{
  name={IPO},
  description={An acronym for \emph{Initial Public Offering}. An IPO is a stock offering of a previously privately-held company for public investment. See also \gls{SEO}}
}
\newglossaryentry{irrational}
{
  name={irrational},
  description={In the theory of financial decision making, a decision maker who is not \gls{vNM.rational}}
}

\newglossaryentry{January.effect}
{
  name={January effect},
  description={In U.S. equity markets, a calendar phenomenon in which abnormal positive returns occur for many stocks at the turn of a year. Prior to its discovery, its effect on small stocks was larger that mid-cap or large cap stocks. After its discovery, it failed to perform as well in succeeding years}
}

\newglossaryentry{kahneman}
{
  name={Daniel Kahneman},
  description={Nobel Laureate in Economics, 2002. The ``K'' in the acronym ``\acrshort{aKnT}''}
}
\newglossaryentry{Karl.Popper}
{
  name={Karl Popper},
  description={Famous philosopher of science, mentor of \gls{George.Soros}},
  sort={Popper}
}
\newglossaryentry{Kelly.Criterion}
{
  name={Kelly Criterion},
  description={A method of determining an investment fraction that maximizes the growth rate of wealth. Also called \gls{Optimal.f}}
}
\newglossaryentry{kernel.smoother}
{
  name={kernel smoother},
  description={A smoothing method that produces an approximating function $\hat{f}(x)$ for a dataset $(x_i,y_i)$ by using a kernel function for weights and a Nadaraya-Watson estimator for $\hat{f}(x)$}
}
\newglossaryentry{k-nearest.neighbors}
{
  name={k-nearest neighbors},
  description={A smoothing method for a scatterplot $x_i,y_i)$ , $i = 1, 2, \ldots, N$ which for a given $x_0$ finds the $k$ nearest $x_i$, and averages the $y_i$ associated with them. See also \gls{loess}}
}
\newglossaryentry{knew.it.all.along.effect}
{
  name={knew-it-all-along effect},
  description={The tendency to recollect the probability of an event after its occurrence as more likely than before its ocurrence. Also known as the \gls{hindsight.bias}}
}
\newglossaryentry{KS-test}
{
  name={KS-test},
  description={The ``KS'' stands for Kolmogorov-Smirnov, the creators of two eponymous tests, (1) the KS one sample test, and (2) the KS two sample test. In the one sample test of sample data coming from a hypothesized distribution $F(x)$, the statistic of interest is 
\[
     D = \underset{x}{max} | \hat{F}(x) - F(x) |, 
\]
  where $\hat{F}(x)$ is the \gls{empirical.distribution.function}. The 2-sample test compares two \glspl{empirical.distribution.function}. Small sample distributions are available for the 1-sample test, whereas only asympotic distributions are available for the 2-sample test}
}

\newglossaryentry{law.of.equal.ignorance}
{
  name={law of equal ignorance},
  description={A probability rule that assigns to each of $N$ possibilities a probability of $1/N$ when very little is known about them. Same as \gls{1/N.rule}}
}
\newglossaryentry{legal.pump.and.dump}
{
  name={legal pump-and-dump},
  description={Legal versions of pump-and-dump, e.g. \gls{strategic.pump.and.dump} and \gls{speculative.pump.and.dump}}
}
\newglossaryentry{leverage}
{
  name={leverage},
  description={Refers to the common financial practice of purchasing or shorting financial instruments by pledging only a fraction of their nominal value. The difference of the cost of the purchase or short is generally supplied at interest from a bank or institution that specializes in lending to trading entities}
}
\newglossaryentry{leveraged.buyout}
{
  name={leveraged buyout},
  description={The takeover of a corporation by \glspl{arbitrageur} who \gls{leverage} the purchase. Part of the \gls{mergersandacquisitions} game}
}
\newglossaryentry{likelihood.function}
{
  name={likelihood function},
  description={Given a sample $x = \{ x_1, x_2, \ldots, x_n \}$ and a family of distributions $F(x ; \theta)$, $\theta \in \Theta$, the function $\mathscr{L}(\theta ; x) = F(x;\theta)$ considered as a function of $\theta$ with $x$ fixed}
}
\newglossaryentry{limit.order}
{
  name={limit order},
  description={An order to buy or sell at a given price (the limit price) or better, optionally with a time to leave the order open. See also \gls{market.order} and \gls{stop.order}}
}
\newglossaryentry{loess}
{
  name={loess},
  description={A smoothing method for a scatterplot $x_i,y_i)$ , $i = 1, 2, \ldots, N$ which for a given $x_0$ finds the $k$ nearest $x_i$ (where $ k = \alpha N$), and averages the $y_i$ associated with them. See also \gls{k-nearest.neighbors}}
}
\newglossaryentry{LOR}
{
  name={Leland O'Brien Rubinstein Associates},
  description={The firm whose principals developed, advocated and marketed \gls{portfolio.insurance} prior to the \acrlong{aWSMC87}. They subsequently admitted the likelihood that portfolio insurance was a major cause of that Crash}
}
\newglossaryentry{log-relative}
{
  name={log-relative},
  description={In finance, given price $p$ at time $t'$ and $q$ at time $t'' > t'$, the log-relative is $log(q/p)$}
}
\newglossaryentry{logarithmic.return}
{
  name={logarithmic return},
  description={log-relative}
}
\newglossaryentry{lognormal.distribution}
{
  name={lognormal distribution},
  description={$Y$ has a lognormal distribution with parameters $\mu$ and $\sigma^2$ if $y \sim exp(X)$, where $X \sim N(\mu,\sigma^2)$ has a Gaussian distribution with mean $\mu$ and variance $\sigma^2$}
}
\newglossaryentry{log.return}
{
  name={log return},
  description={A logarithmic return}
}
\newglossaryentry{long.memory}
{
  name={long memory},
  description={A technical term for a stochastic process $X_t$ that requires its ``correlations to die off slowly.'' One definition requires that for each time $t$, the sum of correlations $\sum_{s=1}^{\infty} \rho_{t+s}$ diverges}
}
\newglossaryentry{long-tailed}
{
  name={long-tailed},
  description={Referring to a distribution with the property
  \[
     \underset{ \rightarrow \infty}{lim} \frac{P \left[ X > x + y \right]}{P \left[ X > x \right]} = 1
  \]
  }
}
\newglossaryentry{lookahead.bias}
{
  name={lookahead bias},
  description={In trading system development, the (often inadvertent) development and testing of a trading system using information assumed to be known at trade time, but in actuality known only at some future time. Some real life examples: (1) using a sentiment index dated on Friday of each week, but not made available until the following Tuesday, and (2) simulating a trade in the middle of a period based on the mean price of the entire period, but then in practice using the mean available from the beginning of the period to the trade time}
}
\newglossaryentry{lookback.horizon}
{
  name={lookback horizon},
  description={Trading system jargon for the period of history needed to make a trading decision. For example, if market history for the past week is all that is needed to decide to buy, sell or do nothing, then the lookback horizon is one week}
}
\newglossaryentry{loss.aversion}
{
  name={loss aversion},
  description={A bias in which people attempt to avoid losses, even if such action gives away considerable \gls{edge}}
}
\newglossaryentry{lottery}
{
  name={lottery},
  description={In \gls{utility.theory}, a gamble that returns one and only one prize $A_1, A_2, \ldots, A_n$ with known probabilities $p_1, p_2, \ldots, p_n$. Notation: 
\begin{equation*}
  ( p_1,A_1; \; p_2, A_2, \; \ldots, \; p_n, A_n )
\end{equation*}
}}
\newglossaryentry{LTCM}
{
  name={Long Term Capital Management},
  description={A \gls{hedge.fund} formed in $1994$ by ex-Solomon Brothers ``whiz kids,'' featuring in addition two Nobel Laureates in Economics, who had developed and advocated the ``new finance \gls{arbitrage} strategies'' that it allegedly used. After its notorous collapse in $1998$, the Federal Reserve, fearful of a run on shadow banking, engineered a rescue by a consortium of $14$ banks. A clear demonstration of the Federal Reserve's real mandate, not the sanitized official version}
}

\newglossaryentry{Manager.of.Managers}
{
  name={Manager of Managers},
  description={In the nonstandard usage of this work, a investment fund that does not itself trade its funds, but hires other managers for that purpose. See also \gls{Trader/Manager}}
}
\newglossaryentry{market.capacity}
{
  name={market capacity},
  description={An informal concept which expresses the idea that fundamental values in a security cannot be maintained when the demand for the security too large. When the demand greatly exceeds the supply, the market can become severly \emph{\gls{overbought}} leading to the likelihood of a correction later}
}
\newglossaryentry{market.ecology}
{
  name={market ecology},
  description={A market ecology at a moment in time includes both the \gls{strategic.ecology} and market structure, in the sense of a game that has rules, venues and technology within which strategies can be enacted. Also called \gls{ecology}}
}
\newglossaryentry{market.fundamentalism}
{
  name={market fundamentalism},
  description={A theory decocting matters of great financial complexity into a simplistic recipe --- that matters of greatest importance to societal welfare are always best served in a laissez-faire stew}
}
\newglossaryentry{market.maker}
{
  name={market maker},
  description={A trader who specializes in providing liquidity to markets by buying and selling to counterparties at either posted or negotiated prices. Market makers are usually exchange-sanctioned and pay low fees to trade. In \glspl{continuous.auction.market}, market makers post bids and offers that the public can trade against. Market makers perform well in non-volatile markets, but tend to withdraw bids and offers in volatile ones and thereby, may exacerbate large price moves}
}
\newglossaryentry{market.order}
{
  name={market order},
  description={An order to buy or sell at the best available price. See also \gls{limit.order} and \gls{stop.order}}
}
\newglossaryentry{Market.Neutral.Strategy}
{
  name={Market Neutral Strategy},
  description={A \emph{Market Neutral Strategy (\acrshort{aMNS})} is any strategy that selects from a universe of assets a portfolio \(P\) such that the \(\beta\) of \(P\) with respect to that universe is nearly zero},
  plural={Market Neutral Strategies}
}
\newglossaryentry{market.paradigm}
{
  name={market paradigm},
  description={A collection of beliefs about the current state of the market that leads to one of three judgments: the market is (1) undervalued, (2) overvalued or (3) neither (1) nor (2)},
}
\newglossaryentry{market.sentiment}
{
  name={market sentiment},
  description={A term that refers to the range of opinions and attitudes of the players in a market. In financial markets, diverse attitudes are often simplified to: bullish, bearish and neutral; in real estate, to prices ``too high'', ``too low'' or ``afforable.'' Sentiment has been believed to be important in selecting market tops and bottoms --- low bullish sentiment predicting an advancing market, and high bullish sentiment predicting a declining market. Same as \gls{sentiment}}
}
\newglossaryentry{market.tone}
{
  name={market tone},
  description={A term that refers to the way that a market reacts to positive and negative news. A market has good tone if it has a positive reaction to good news and negative reaction to negative news. It has poor tone if it has no or a negative reaction to good news and a negative reaction to negative news. \emph{Market tone} is a soft, rather subjective \gls{sentiment.indicator}}
}
\newglossaryentry{martingale}
{
  name={martingale},
  description={In this work, a time series $X_t$ for which $E[X_t] < \infty$ and $E[X_{t+1} \, | \, X_t ] = X_t$ for all meaningful $t$}
}
\newglossaryentry{martingale.difference.series}
{
  name={martingale difference series},
  description={In this work, a time series $X_t$ is a martingale difference series with respect to \glspl{information.set} $\{\mathcal{I}_t\}$ if for all meaningful $t$, $E[X_t \, | \, \mathcal{I}_{t-1}]  < \infty$ and $E[X_{t} \, | \, \mathcal{I}_{t-1} ] = 0$. When information sets are not specified, it is understood that each $\mathcal{I}_t = \{X_t, X_{t-1}, X_{t-2}, \ldots\}$ is the history of the series up to time $t$}
}
\newglossaryentry{maximum.likelihood}
{
  name={maximum likelihood},
  description={A method of producing estimators from a sample $x = \{ x_1, x_2, \ldots, x_n \}$ assmuing that the data arise from a family of distributions $F(\theta)$ indexed by $\theta \in \Theta$. The \gls{likelihood.function} $\mathscr{L}(\theta ; x) = F(x;\theta)$ is considered to be a function of $\theta$ with the sample $x$ fixed. The \gls{maximum.likelihood.estimator} is the distribution(s) $F(\theta_{max}$ at which the \gls{likelihood.function} is maximized as a function of $\theta$}
}
\newglossaryentry{maximum.likelihood.estimator}
{
  name={maximum likelihood estimator},
  description={Given a sample $x = \{ x_1, x_2, \ldots, x_n \}$ and a family of distributions $F(x ; \theta)$, $\theta \in \Theta$, the distribution $F(\theta_{max})$ for which the \gls{likelihood.function} $\mathscr{L}(\theta ; x) = F(x;\theta)$ is maximized as a function of $\theta$}
}
\newglossaryentry{mccauley}
{
  name={Joseph L. McCauley},
  description={Well-known specialist in chaos theory and econophysics}
}
\newglossaryentry{Mean.Variance.Portfolio.Theory}
{
  name={Mean-Variance Portfolio Theory},
  description={A theory that develops a notion of efficient portfolios as ones belonging to an efficient frontier. The development requires only the means and covariances of instruments in a portfolio}
}
\newglossaryentry{mean-variance.world}
{
  name={Mean-Variance World},
  description={A fictitious world featuring the curious analysis of complex stochastic data assuming that only the first two moments of the distributions are necessary. Inhabited only by economists and their lackeys}
}
\newglossaryentry{mechanical.trading.system}
{
  name={mechanical trading system},
  description={In trading and investment, a collection of well-defined rules that specify when to enter the market, the size and side (buy or sell) for entry, and when to exit all or part of an existing position. In this work, the same as a \gls{trading.system}}
}
\newglossaryentry{mental.accounting}
{
  name={mental accounting},
  description={A compartmentalization of financial activity into \emph{mental acounts} in which gains or losses in any one account are not aggregated with gains or losses in the others}
}
\newglossaryentry{mergersandacquisitions}
{
  name={mergers \& acquisitions},
  description={The investment banking and large funds business of corporate restructuring by mergers and \glspl{leveraged.buyout}}
}
\newglossaryentry{Miller's.paradox}
{
  name={Miller's paradox},
  description={An \acrshort{aEMH} paradox due to Edward Miller, 1977. He applied \gls{winner's.curse} arguments to market auctions and concluded, inter alia, that (1) prices will not be fair and investment capital will be inefficiently allocated, and (2) the riskiest stocks will be the most overpriced. These conclusions are so embarrassing to EMH theory that they are seldom written about by academic economists}
}
\newglossaryentry{minskymoment}
{
  name={minsky moment},
  description={A time in the business cycle at which financial values suddenly collapse}
}
\newglossaryentry{mixed.strategy}
{
  name={mixed strategy},
  description={In \gls{game.theory}, a strategy in which each player independently chooses a strategy using a private probability distribution. Also known as a \gls{randomized.strategy}},
  plural={mixed strategies}
}
\newglossaryentry{momentum}
{
  name={momentum},
  description={In academic finance, a \gls{statistical.regularity} in which a portfolio that has appreciated has the tendency to continue appreciating, and a portfolio that has depreciated has a tendency to continue to depreciate. Note that this usage does not correspond directly to ``momentum'' in technical analysis, but rather to ``\glslink{trend}{trending}.'' The term came into widespread use following the discoveries by Jegadeesh \& Titman \cite{citeulike:940976} that the top decile of stocks ranked on 1-year returns contined to appreciate for about another year, and the bottom decile continued to drop for another year. In \gls{technical.analysis}, called \glslink{trend}{trending}}
}
\newglossaryentry{Momentum.Puzzle}
{
  name={Momentum Puzzle},
  description={The findings of two \acrlong{aJnT} famous studies, \cite{citeulike:940976} (1993) and \cite{citeulike:4057155} (2001) showed convincingly that the \gls{efficient.market.hypothesis} \acrshort{aEMH} is untenable. It destroyed all three pillars of the \acrshort{aEMH} by exhibiting that a simple \gls{momentum} strategy produces excess returns, and that \gls{arbitrage} between $1993$ and $2001$ failed to eliminate it. The puzzle is
\begin{enumerate}
\item Why do U.S. stock returns that have recent \gls{momentum} tend to persist over periods averaging a year?
\item Why do U.S. stock returns over $3$ years tend to revert over the next $3$ years?
\item Why did \gls{arbitrage} fail to eliminate the \gls{momentum} anomaly after the publication of the first \acrshort{aJnT} study, i.e. between $1993$ and $2001$?
\end{enumerate} }
}
\newglossaryentry{moving.average}
{
  name={moving average},
  description={A moving average is a time series method of averaging past observations, usually for purposes of prediction. For a univariate or multivariabe time series $\{X_t\}$, a generic moving average has the form $MA_t = \sum_{s=0}^{s=\infty} \alpha_s X_{t-s}$ where $\alpha_s$ are fixed weights for which $\sum \alpha_s = 1$. There are various types of moving average: \gls{simple.moving.average}, \gls{weighted.moving.average} and \gls{exponential.moving.average}. See glossary entries for definitions of these types}
}
\newglossaryentry{Muller-Lyer.Illusion}
{
  name={M\"uller-Lyer Illusion},
  description={A famous visual illusion from Gestalt psychology in which two lines of equal length, one with inward-pointing arrows at the ends, the other with outward-pointing arrows, are perceived as having different lengths}
}
\newglossaryentry{musical.chairs}
{
  name={musical chairs},
  description={A popular child's game in which $n$ children walk in a circle with $n-1$ chairs close by while music plays. When the music stops, each rushes to sit in a chair, leaving one hapless child without a chair. That child is eliminated, and the game continues, always with one chair less than the number of children, until some prescribed number is left. Also reported to be played by some (presumably feeble-minded) adults. A variant occurs in the \gls{DSSW.Model}: (1) the music stops permanently at some random point, and (2) the goal is not to win, but to be eliminated as close to the end as possible}
}
\newglossaryentry{mutation}
{
  name={mutation},
  description={A process whereby some parent genes are copied inexactly}
}
\newglossaryentry{mutual.fund}
{
  name={mutual fund},
  description={A mutual fund is an investment company that sells shares in a portfolio it manages using the pooled monies from investors. In the United States, mutual funds are registered with the Securities and Exchange Commission and are regulated by the Investment Company Act of $1940$. There are three types of mutual funds in the United States: open-end funds, closed-end funds and unit investment trusts. Open-end funds are publicly traded and are required to offer new shares and redeem existing ones at certain times only, usually at the end of each business day. Closed end funds at creation offer shares which are not redeemable until the fund either liquidates or converts to an open end fund. Unit investment trusts form fixed portfolios at creation, with a specific termination date, and in a public offering sell shares to the public. While investors in closed end funds and unit investment trusts cannot redeem their shares prior to ending, secondary markets often allow investors to sell shares prior to that time. Mutual funds may be either actively or passively managed}
}
\newglossaryentry{myopically.rational}
{
  name={myopically rational},
  description={Referring to na\"{i}ve, putatively self-interested strategies pursued by many investors. Such strategies can cause major market inefficiencies such as bubbles and crashes},
}

\newglossaryentry{Nash.equilibrium}
{
  name={Nash equilibrium},
  description={A concept introduced by John Nash in $1950$ in which players' strategies cannot be unilaterally improved upon. Nash showed that under general conditions, such strategies always exist. See Myerson, \cite{myerson1997game} for a presentation},
  plural={Nash equilibria}
}
\newglossaryentry{noisy.channel.model}
{
  name={Noisy Channel Model},
  description={A model (the \acrshort{ancm}) due to Martin Hilbert that explains eight behavioral biases using a parsimonious model based on Claude \glslink{shannon}{Shannon's} information theory}
}
\newglossaryentry{negative.sum}
{
  name={negative-sum},
  description={A real or artificial mathematical game in which the sum of payoffs to the players is negative. Negative-sum games typically occur because a referee or game organizer charges a fee to play}
}
\newglossaryentry{neuroeconomics}
{
  name={neuroeconomics},
  description={The study of brain activity during economic decision making, such as buying a car or investing in a stock. When only financial instruments are involved, called \emph{neurofinance}}
}
\newglossaryentry{New.Deal}
{
  name={New Deal},
  description={The slogan adopted by U.S. President Franklin Roosevelt for his progressive program during the \gls{Great.Depression}. In financial markets, the New Deal involved regulation of the financial system to punish undue speculation and afford a measure of protection to average investors}
}
\newglossaryentry{no.arbitrage.principle}
{
  name={no-arbitrage principle},
  description={For pure \gls{arbitrage}, the principle that arbitrage cannot yield a sure profit. For risk arbitrage, the principle that arbitrage cannot lead to excess expected risk-adjusted returns. The principle is ordinarily invoked to show that arbitrage cannot lead to consistent profits, which recalls the joke about the economist and the statistician \ldots in Volume 2,}
}
\newglossaryentry{noise.trader}
{
  name={noise trader},
  description={(1) In the \emph{noise trader model of efficient markets}, an investor whose decisions may be treated as random. In this model, all investors are either rational or noise traders. See also \gls{efficient.market.hypothesis} (2) In the \gls{DSSW.Model} of bubbles and crashes, a na\"{i}ve \gls{trend.follower} who is left unseated in that game of \gls{musical.chairs}}
}
\newglossaryentry{nominal.gain}
{
  name={nominal gain},
  description={In terminology developed for this book, the \emph{nominal gain} is the total amount won (or if lost, a negative value) by a trading system when the same amounts are bet at every trade. See also \gls{adjusted.nominal.gain}}
}
\newglossaryentry{nonlinear.regression}
{
  name={nonlinear regression},
  description={A model of the form $y = f(x,e)$, where $y$ is dependent variable, $x$ is an independent variable, $e$ is an ``error'' or ``fluctuation,'' and $f$ is a nonlinear function of $(x,e)$. $x$ and $y$ may be multidimensional vectors}
}
\newglossaryentry{nonparametric.model}
{
  name={nonparametric model},
  description={In statistics, a \emph{nonparametric model} is one that makes no explicit assumption about a finite-dimensional family of distributions. Some examples are nonparametric regression, quantile regression, generalized additive models such as \acrshort{aACE}, \acrshort{aAVAS} and \acrshort{aGAM}}
}
\newglossaryentry{normal.form}
{
  name={normal form},
  description={An \gls{artificial.mathematical.game} in which the \glspl{strategy} available to the players are indicated by elements of a set $C$. This set in simple cases is finite, but in most games is infinite, and may represent a complete specification of every choice a player might make in some game contingency. Also called the \gls{strategic.form}}
}
\newglossaryentry{normative.decision.science}
{
  name={normative decision science},
  description={See \gls{decision.science}}
}
\newglossaryentry{normative.to.descriptive.step}
{
  name={normative-to-descriptive step},
  description={The requirement that studies be performed to verify that a normative economic model adequately describes real economic behavior (Michael Bacharach \cite{bacharach1977economics})}
}

\newglossaryentry{observational.data}
{
  name={observational data},
  description={Data acquired through observation without the use of experimental methods}
}
\newglossaryentry{observational.experimental.stage}
{
  name={observational/experimental stage},
  description={The stage of scientific inquiry that collects experimental data}
}
\newglossaryentry{Optimal.f}
{
  name={Optimal $f$},
  description={A method of determining an investment fraction that maximizes the growth rate of wealth. Also called the \gls{Kelly.Criterion}}
}
\newglossaryentry{order}
{
  name={order},
  description={In trading, an \emph{order} is a request, possibly contingent, to buy or sell a financial instrument. \emph{Orders} are time-stamped requests that specify quantity and contingently, price limits, immediate execution, cancellation if not filled, and so on}
}
\newglossaryentry{Ordinal.Utility}
{
  name={Ordinal Utility},
  description={See \gls{utility.theory}}
}
\newglossaryentry{outlier}
{
  name={outlier},
  description={A unusually large or small obsevation in a data sample}
}
\newglossaryentry{outlier-proneness}
{
  name={outlier-proneness},
  description={In dynamic systems, the tendency to produce outliers}
}
\newglossaryentry{outstanding.shares}
{
  name={outstanding shares},
  description={The number of shares issued by a company, which includes restricted shares and the \gls{float}}
}
\newglossaryentry{overbought}
{
  name={overbought},
  description={Said of a financial instrument that is priced too high due excessive ``buying pressure.'' See also \gls{oversold}}
}
\newglossaryentry{overreaction}
{
  name={overreaction},
  description={In financial markets, overreaction refers to an excessively large price move in response to an event. See also \gls{underreaction}}
}
\newglossaryentry{oversold}
{
  name={oversold},
  description={Said of a financial instrument that is priced too low due to excessive ``selling pressure.'' See also \gls{overbought}}
}

\newglossaryentry{pairs.trading.strategy}
{
  name={pairs trading strategy},
  description={A \emph{pairs trading strategy (PTS)} is any strategy that selects from some universe of stocks those price-pairs that have a putatively overvalued and undervalued member, and nearly simultaneously buy the undervalued and  sell the overvalued. The pair so-formed is then liquidated when prices are judged to return to ``equilibrium.''},
  plural={pairs trading strategies}
}
\newglossaryentry{parametric.model}
{
  name={parametric model},
  description={In statistics, a model in which the data are generated by a distribution drawn from a family of distributions, $\{ F_{\theta} \, | \, \theta \in \Theta \}$, where $\Theta$ is a finite dimensional parameter space. Estimation in parametric models requires finding the parameter $\theta_0 \in \Theta$ that in some sense provides a best fit for observed data. Some standard fitting approaches are \gls{maximum.likelihood}, Bayesian methods and least squares estimation with normally distributed errors}
}
\newglossaryentry{pareto}
{
  name={Villfredo Pareto},
  description={Italian engineer, sociologist, economist, political scientist and philosopher who was one of the first to apply scientific analysis to markets. (1848~--~1923)}
}
\newglossaryentry{Pareto.efficient}
{
  name={Pareto efficient},
  description={In decision or game theory, a \gls{strategy.profile} $s$ is Pareto efficient iff given players $N$, payoff functions $u_i$, $i \in N$, there is no $i \in N$ and \gls{strategy.profile} $s'$ such that $u_i(s') > u_i(s)$ unless there is a $j \in N$ such that $u_j(s) < u_j(s')$},
  plural={Pareto efficiency}
}
\newglossaryentry{partial.disclosure}
{
  name={partial disclosure},
  description={A non-standard term in \gls{game.theory} in which a player discloses that he or she will pursue a certain strategy with some probability $p$, $0 < p < 1$ and $p \ne 0.5$}
}
\newglossaryentry{peak.to.trough.drawdown}
{
  name={peak-to-trough drawdown},
  description={In a financial time series, the largest loss from a previous high}
}
\newglossaryentry{percentage.return}
{
  name={percentage return},
  description={In finance, given price $p$ at time $t'$ and $q$ at time $t'' > t'$, the percentage return is $100*(q-p)/p$}
}
\newglossaryentry{perfect.memory}
{
  name={perfect memory},
  description={A time series $\{X_t\}$ with information sets $\{I_t\}$ has perfect memory if $I_t \supseteq \{ X_{t}, X_{t-1}, \ldots \}$}
}
\newglossaryentry{personal.probability}
{
  name={personal probability},
  description={A theory of probability in which each person makes his or her own subjective, consistent probability estimates. Sometimes called \emph{subjective probability}}
}
\newglossaryentry{pick.the.best.heuristic}
{
  name={pick-the-best heuristic},
  description={A \gls{Fast.Frugal.Heuristics} that makes decisions by examining cues for two choices until the first one that is better is found}
}
\newglossaryentry{POPP.trend}
{
  name={POPP trend},
  description={A trend that occurs as part of a \gls{Pursuit.of.Profits.Paradigm} (POPP)}
}
\newglossaryentry{portfolio.insurance}
{
  name={portfolio insurance},
  description={A strategy for insuring against portfolio losses by dynamical replication of a put on a benchmark index. Popular before the worldwide stock market crash of $1987$, but not subsequently due to its poor performance during that crash}
}
\newglossaryentry{position}
{
  name={position},
  description={In a trading context, either a \emph{long position}, which consists of ownership of an instrument or a fungible contract used in trading, or a \emph{short position} that consists of the opposite of a long position (except for financing costs), created synthetically through stock borrowing or directly through ownership of the opposite side of a long contract position. Same as a \gls{trading.position}}
}
\newglossaryentry{positional.format}
{
  name={positional format},
  description={The representation of a trading history as time-stamped cumulative positions}
}
\newglossaryentry{positive.affine.transformation}
{
  name={positive affine transformation},
  description={A positive affine transformation of $u \in \mathbb{R}$ is a function $g(u) = a u \, + \, b$ where $a,b \in \mathbb{R}$ are constants and $a > 0$}
}
\newglossaryentry{positive.feedback}
{
  name={positive feedback},
  description={In a financial context, a process in which perturbations in the process are amplified as it progresses, which leads to system instability. Negative feedback processes damp perturbations, leading to stability}
}
\newglossaryentry{post.event.price.drift}
{
  name={post-event price drift},
  description={A market anomaly in which an event's positive or negative surprises lead to price drift up or down thereater. The best known example is the \emph{post earnings announcement drift} in which the averge drift of prices after the announcement is strictly increasing in the \gls{SUE} (Standardized Unexpected Earnings) of the announcement}
}
\newglossaryentry{potentially.profitable.gambling.system}
{
  name={potentially profitable gambling system},
  description={A potentially profitable gambling system (PPGS) is a betting system for a financial time series $\{X_t\}$ that has at least one positive expected return},
} 
\newglossaryentry{power.law}
{
  name={power law},
  description={A distribution concentrated on $(0,\infty)$ having a cumulative distribution of the form $F(x) = 1 - x^{-\alpha}$, where $\alpha > 0$. $\alpha$ is called the \emph{exponent} of the power law}
}
\newglossaryentry{power.law.tail}
{
  name={power law tail},
  description={A distribution whose upper and/or lower tails are asymptotically equivalent to $(x/x_0)^{-\alpha}$, where $x_0 > 0$ is a constant and $\alpha > 0$. As on \glspl{power.law}, $\alpha$ is called the \emph{exponent} of the power law tail}
}
\newglossaryentry{PPGS.renewal.process}
{
  name={PPGS renewal process},
  description={A potentially profitable gambling system renewal process is an algorithm that (1) produces transactions depending on the state of the market and its own trading history, (2) as a function of those inputs, produces trades each consisting of a founding transaction that initiates a position from a state having no position, a closing transaction that liquidates remaining positions leaving a zero position, and has no other transactions between founding and closing that change the founding position from long to short or short to long, and (3) produces a potentially unending stream of trades (renewal)},
  plural={PPGS renewal processes}
}
\newglossaryentry{precision}
{
  name={precision},
  description={In the theory of statistical estimation, an estimator is called \emph{precise} if its standard deviation is small and otherwise is called \emph{imprecise}}
}
\newglossaryentry{predictability}
{
  name={predictability},
  description={In a dynamic systems, the propensity to forecast its future evolution, if only stochastially. A system can be stongly predictable or weakly predictable}
}
\newglossaryentry{preference.order}
{
  name={preference order},
  description={A ranking among a set $\Omega$ of events that satisfies three axioms: \gls{Complete.Ordering}, \gls{Reflexivity.Of.Preference} and \gls{Transitivity}}
}
\newglossaryentry{preference.reversal}
{
  name={preference reversal},
  description={A \gls{decision.problem} in which the independence of irrelevant alternatives is violated so that $A \succ B$ when $C$ is present, and $B \succ A$ if $C$ is absent}
}
\newglossaryentry{present.value}
{
  name={present value},
  description={The value of a stream of future cash payments using the method of discounting with a rate schedule}
}
\newglossaryentry{price.distorter}
{
  name={price distorter},
  description={Any exogenous news, constraints on trading, widely held beliefs, or widely used strategies that effect \glspl{price.impact}. For example, the fact that mutual funds cannot short stocks suggests that prices might be higher than they would be if shorting were allowed. Same as \gls{distortion.factor}}
}
\newglossaryentry{price.impact}
{
  name={price impact},
  description={The change in price due to a trade or sequence of trades, as a function of price, quantity and market depth. Note that concept is ill-defined, in the sense that it assumes a counterfactual price that would have obtained in the absence of that trade or trade sequence. Since markets are replete with ``nuisance variables'' that are demonstrably important in determining price impact, in practice one usually settles for estimates of average impact using statistical models}
}
\newglossaryentry{Price.Impact.Law}
{
  name={Price Impact Law},
  description={A part of the \glspl{Half.Cubic.Law} which states that the average normalized price impact is approximately proportional to the square root of the average normalized quantity traded}
}
\newglossaryentry{price.resistance}
{
  name={price resistance},
  description={In \gls{technical.analysis}. a price that acts as a barrier to price increases. A stock has resistance at $\$50$, for example, if the price approaches that level, usually twice or more, but each time fails to exceed it. Resistance is strong if it bounces hard off that level, usually with significant size offered at that level, or weak otherwise. See also \gls{resistance}, \gls{price.support} and \gls{support}}
}
\newglossaryentry{price.support}
{
  name={price support},
  description={In \gls{technical.analysis}. a price that acts as a barrier to price decreases. A stock has support at $\$50$, for example, if the price approaches that level, usually twice or more, but each time fails to drop below it. Support is strong if it bounces hard off that level, usually with significant size offered, or weak otherwise. See also \gls{support}, \gls{price.resistance} and \gls{resistance}}
}
\newglossaryentry{priming}
{
  name={priming},
  description={An effect in which the \gls{associative.machine} uses ambient information, including irrelevant information, to make a decision}
}
\newglossaryentry{principal.components.analysis}
{
  name={principal components analysis},
  description={A method of decomposing a $n$-multivariate dataset into $m < n$ mutually orthogonal \glspl{direction} such that the sum of the variances asociated with the directions is maximal over all other sets of $m$ directions. \acrshort{aPCA} results from performing an eigenvalue decomposition of the covariance matrix and selecting eigenvectors associated with the $m$ largest eigenvalues},
  plural={principal components analyses}
}
\newglossaryentry{prisoners.dilemma}
{
  name={prisoner's dilemma},
  description={A famous two player artificial mathematical game in which the unique Nash equilibrium is not Pareto efficient. Often called a paradox, but that is incorrect. It is a consequence of the requirements of non-cooperative play that precludes the players from cooperating}
}
\newglossaryentry{prescriptive.decision.science}
{
  name={prescriptive decision science},
  description={See \gls{decision.science}}
}
\newglossaryentry{problem.of.multiplicity}
{
  name={problem of multiplicity},
  description={If tests are independent, then testing each of $n$ hypotheses at a significance level $p$ results in a probability $1 - (1 - p)^n$ of rejecting at least one, a quantity that approaches 1 as $n \rightarrow \infty$. For example, if 40 independent hypotheses are each tested at a $5\%$ level of significance level, then the probability of at least one rejection is $1 - 0.95^40 = 0.87$}
}
\newglossaryentry{probability.matching}
{
  name={probability matching},
  description={A behavior that occurs in experiments when a reward has probabilities known to experimenters but not to experimental subjects. In such experiments, both animals and humans eventually randomize their choices in alignment with those probabilities. Such behavior is irreconcilable with \acrlong{arct}}
}
\newglossaryentry{probability.space}
{
  name={probability space},
  description={A probability space $\mathcal{U}$ consists of (1) a set of \emph{outcomes} $\Omega$,
(2) a sigma-field $\mathcal{E}$ of \emph{events} on $\Omega$, and (3) a \emph{probability} function $P: \mathcal{E} \rightarrow [0,1]$ for which (i) $P[\Omega] = 1$, and (ii) for any disjoint family $\{E_i\}_{i=1}^{i=\infty} \subset \mathcal{E}$, $P[\bigcup_{i=1}^{i=\infty}] = \sum_{i=1}^{i=\infty} P[E_i]$}
}
\newglossaryentry{professional.gamblers}
{
  name={professional gamblers},
  description={A person whose livelihood consists of enlightened risk taking. In the economic theory of markets, the set of all professional gamblers in financial markets is empty}
}
\newglossaryentry{profile}
{
  name={profile},
  description={In the technical trading terminology ot this book, a \emph{profile} or \gls{signal.profile} is a collection of vectors $v_t \in \mathbb{R}^n$, that at each time $t$ gives a trading signal for trading variables $p_t \in  \mathbb{R}^n$ as a scalar product $v_t \cdot p_t$. If $v_t = v$ is constant for all $t$, then the \emph{profile} is called \emph{stationary}. For the most part, methods used in this book produce stationary profiles}
}
\newglossaryentry{profit.and.loss}
{
  name={profit \& loss},
  description={and abbreviation for \emph{profit and loss}},
  plural={profits \& losses}
}
\newglossaryentry{prospect}
{
  name={prospect},
  description={A lottery}
}
\newglossaryentry{prospect.theory}
{
  name={prospect theory},
  description={A descriptive model of human decision making in risky and uncertain situations}
}
\newglossaryentry{pure.arbitrage}
{
  name={pure arbitrage},
  description={An \gls{arbitrage} having such small risk that a profit is almost certain, e.g. the near simultaneous purchase and sale of the same security on two different exchanges at advantageously different prices}
}
\newglossaryentry{pure.strategy}
{
  name={pure strategy},
  description={In \gls{game.theory}, a strategy in a player's strategy set},
  plural={pure strategies}
}
\newglossaryentry{pure.strategy.profile}
{
  name={pure strategy profile},
  description={In \gls{game.theory}, a strategy in the Cartesian product of players' strategy sets}
}
\newglossaryentry{Pursuit.of.Profits.Paradigm}
{
  name={Pursuit of Profits Paradigm},
  description={A type of \gls{strategic.evolution} in which a ``chase after riches'' leads to boom-bust cycles in markets}
}

\newglossaryentry{random.variable}
{
  name={random variable},
  description={A mapping from a \gls{probability.space} to a space of values. Intuitively, a varying quantity which may be thought of as arising from an underlying and unseen probabilistic process. A real-valued random variable is one whose values are real numbers}
}
\newglossaryentry{random.walk}
{
  name={random walk},
  description={A time series $X_t$, $t \in \mathbb{N}$ such that $X_t = \mu + X_{t-1} + U_t$, where $\{U_t\}$ is an \gls{i.i.d.white.noise}}
}
\newglossaryentry{randomized.game}
{
  name={randomized game},
  description={In \gls{game.theory}, a game which is derived from a basic game by extending strategies to randomized strategies and payoffs to expected utilities of \glspl{strategy.profile}}
}
\newglossaryentry{randomized.strategy}
{
  name={randomized strategy},
  description={In \gls{game.theory}, a strategy in which each player independently chooses a strategy with a known private probability distribution. Also known as a \gls{mixed.strategy}},
  plural={randomized strategies}
}
\newglossaryentry{rational}
{
  name={rational},
  description={Referring to a being in an enlightened state of \gls{rationality}. It is an open question, however, as to whether a being of such intellect would choose to operate as if guided by a \gls{utility.function}}
}
\newglossaryentry{rational.actor}
{
  name={rational actor},
  description={A decision maker who applies a von Neumann/Morgenstern (vNM) utility function to \glspl{decision.problem}}
}
\newglossaryentry{rational.choice.theory}
{
  name={rational choice theory},
  description={The astounding proposal that everybody acts at each and every waking moment as though they seek the most cost-effective way to achieve their goals, irrespective of the worthiness of those goals --- definitely good for rapacious psychopaths, bad for everybody else}
}
\newglossaryentry{rational.expectations}
{
  name={rational expectations},
  description={The curious proposal that the average human is a vNM automaton, or failing that, that an average human can be treated as a vNM automaton, or if that be unacceptable, that average groups of humans act as though they are vNM automatons, and in the event that this is found wanting, that these assumptions are close enough for economic work \ldots justifying the momentous conclusion that the best guess about the future is people's average opinion today}
}
\newglossaryentry{rational.speculator}
{
  name={rational speculator},
  description={In the \gls{DSSW.Model} of bubbles, a trader who acts with other like-minded traders to continue \glspl{trend} beyond an assets' fundamental values. These traders need to be nimble in order to exit before price corrects --- essentially a game of \gls{musical.chairs}. This strategy can be called \gls{legal.pump.and.dump}}
}
\newglossaryentry{rationality}
{
  name={rationality},
  description={Referring to an elightened state of being characterized by making all, or at least all financial, decisions as if guided by a \gls{utility.function}}
}
\newglossaryentry{R.language}
{
  name={R Statistical Language},
  description={A statistical and data analysis software package available at URL: \url{www.r-project.org}}
}
\newglossaryentry{reading.people}
{
  name={reading people},
  description={In gambling, the art of predicting from a person's gestures, mannerisms and expressions the mistakes they're about to make}
}
\newglossaryentry{real.game}
{
  name={real game},
  description={A real stuctured contest between several persons who engage strategically to determine their payoffs. See also \gls{game.theory} and \gls{artificial.mathematical.game}}
}
\newglossaryentry{rebalancing}
{
  name={rebalancing},
  description={In finance, the act of modifying portfolio allocations by trading only the differences between the current and target allocaions. All public funds perform periodic rebalancing to provide for addition or withdrawl of invested funds or to modify the current portfolio}
}
\newglossaryentry{recency}
{
  name={recency},
  description={In psychology, the strong tendency to give more recent events greater weight in decisions. Is important in the \Gls{availability} Heuristic}
}
\newglossaryentry{recognition.heuristic}
{
  name={recognition heuristic},
  description={A \gls{Fast.Frugal.Heuristics} that selects from a menu of alternatives the choice that is most ``recognizable''}
}
\newglossaryentry{Reflexivity.Of.Preference}
{
  name={Reflexivity of Preference},
  description={An axiom of utility theory that requires for each event $A$, $A \succeq A$}
}
\newglossaryentry{reflexivity}
{
  name={reflexivity},
  description={A philosophical framework for understanding social phenomena, but especially financial markets, developed by hedge fund manager \gls{George.Soros}}
}
\newglossaryentry{regret}
{
  name={regret},
  description={\emph{Regret} is used in two senses in this work: (1) as the feeling or emotion of sorrow for a past action or choice, and (2) as a rule in a \gls{decision.problem} that selects the choice which minimizes the probability of worst loss}
}
\newglossaryentry{relative.value.strategy}
{
  name={relative value strategy},
  description={Any strategy that shorts an ``overvalued'' portfolio and buys and ``undervalued'' one. The entire portfolio can be \emph{market neutral}, meaning that it has little expected gain or loss when the market moves. It can also be \emph{dollar neutral}, meaning that it requires a net zero investment ignoring fees},
  plural={relative value strategies}
}
\newglossaryentry{representativeness}
{
  name={representativeness},
  description={A decision heuristic that makes the most representative choice}
}
\newglossaryentry{resistance}
{
  name={resistance},
  description={Same as \gls{price.resistance}. See also \gls{price.support} and \gls{support}}
}
\newglossaryentry{return}
{
  name={return},
  description={In finance, a measure of the gain or loss of an investment that does not depend on price. Given an investment purchases at price $p$ and later valued at $q$, the fractional return over that holding period $(q-p)/p$, the percentage return is $100*(q-p)/p$, and the logarithmic return is $log(q/p)$}
}
\newglossaryentry{riding.the.yield.curve}
{
  name={riding-the-yield-curve},
  description={A type of \gls{carry.trade} in which short maturity notes are sold and long maturity ones purchased, and as short maturity ones mature are replaced by new ones. When the yield curve is upward sloping, the trade has positive returns; when it inverts, though, the trade loses}
}
\newglossaryentry{risk}
{
  name={uncertainty},
  description={In financial decision theory, referring to situations whose non-deterministic events are known with known probabilities. See also \gls{uncertainty}}
}
\newglossaryentry{risk-adjusted.return}
{
  name={risk-adjusted return},
  description={An asset return that is ``adjusted'' for risk. The most common method consists of dividing an asset's excess return (the difference of the return and the risk-free rate) by the asset's standard deviation (Sharpe ratio), but other methods of risk-adjusting also exist: the Treynor ratio, the Sortino Ratio, the Sterling Ratio and Jensen's Alpha}
}
\newglossaryentry{risk-free.rate}
{
  name={risk-free rate},
  description={A default-free rate of interest}
}
\newglossaryentry{risky.decision.problem}
{
  name={risky decision problem},
  description={A \gls{decision.problem} that involves only events with known probabilities. See also \gls{uncertain.decision.problem}},
  plural={Risky Decision Problem}
}
\newglossaryentry{risk.arbitrage}
{
  name={risk arbitrage},
  description={Statistical \gls{arbitrage}. See also \emph{pure arbitrage}}
}
\newglossaryentry{risk.averse} 
{
  name={risk averse},
  description={An individual is risk averse at $x$ if their utility function's absolute or relative coefficient of risk aversion at $x$ is positive, $r(x) > 0$. He or she is risk averse if $r(x) > 0$ for all $x$}
}
\newglossaryentry{risk.neutral}
{
  name={risk neutral},
  description={An individual is risk neutral at $x$ if their utility function's absolute or relative coefficient of risk aversion at $x$ is zero, $r(x) = 0$. He or she is risk neutral if $r(x) = 0$ for all $x$}
}
\newglossaryentry{risk.seeking}
{
  name={risk seeking},
  description={An individual is risk seeking at $x$ if their utility function's absolute or relative coefficient of risk aversion at $x$ is negative, $r(x) < 0$. He or she is risk seeking if $r(x) < 0$ for all $x$}
}
\newglossaryentry{robust-yet-fragile}
{
  name={robust-yet-fragile},
  description={A property financial networks in which small shocks have low probabilities of turning into cascades (\gls{robustness}) but there is a small chance of very large cascade (\glslink{fragile}{fragility})}
}
\newglossaryentry{robustness}
{
  name={robustness},
  description={The property of a dynamic system in which the probability of large and rapid changes have very low probability. Not \gls{fragile}}
}
\newglossaryentry{roehner}
{
  name={Bertrand M. Roehner},
  description={Prominent early contributor (physicist) to econophysics}
}
\newglossaryentry{Royal.Dutch.Company}
{
  name={Royal Dutch Company},
  description={A petroleum company headquarted in London, but which shares all profits with the Dutch company, \gls{Shell}. Deviation of the prices of these two companies is the source of a serious EMH \gls{anomaly}}
}
\newglossaryentry{rule.based.perception}
{
  name={rule-based perception},
  description={Perception using rules which generally involves decomposing scenarios into components. As opposed to \gls{holistic.perception}}
}
\newglossaryentry{rule.of.seemingly.unrelated.accidents}
{
  name={rule of seemingly unrelated accidents},
  description={A rule by which the occurrence of frequent, sporadic, seemingly unrelated ``accidents'' in markets augur a crisis, the mecahism being some hidden contagion-provoking connections among apparently unrelated segments of the markets}
}

\newglossaryentry{salience}
{
  name={salience},
  description={In psychology, the strong tendency to give greater weight to observations that for some reason ``stand out'' from others. For example, a red dot in a field of black ones draws attention to itself and is remembered better than the black ones, a traumatic event will be remembered easily and a moving object will stand out against a stationary background. Is important in the \Gls{availability} Heuristic}
}
\newglossaryentry{SAFM}
{
  name={SAFM},
  description={An acronym for the \emph{Strategic Analysis of Financial Markets} framework}
}
\newglossaryentry{satisficing}
{
  name={satisficing},
  description={A term originated by \gls{simon} for \gls{bounded.rationality}. A boundedly rational decision maker, unlike a Superhuman, has limited time and resources, so Simon suggested that a search would consider choices only until finding one that is ``good enough''; this process of making a choice is called \emph{satisficing}}
}
\newglossaryentry{save.more.tomorrow}
{
  name={Save More Tomorrow},
  description={A savings plan designed by Shlomo Benarzi and Richard Thaler based on behavioral principles, that offers employees a program that commits only raises to contributions and changes the default choice to the Save More Tomorrow plan}
}
\newglossaryentry{security.level}
{
  name={security level},
  description={A minimum acceptable wealth level in the \gls{spa} theory}
}
\newglossaryentry{security.potential.criterion}
{
  name={security-potential criterion}, 
  description={In the \gls{spa} theory a decumulatively weighted value rule of the form $SP = \sum_i h(D_i)(v_i - v_{i-1})$, where the decumulative probabilities are $D_i = \sum_{j=i}^{i=n} p_j$, and the function $h$ has the form $h(D) = w D^{q_s + 1} + (1-w)(1 - (1 - D)^{q_p + 1}$}
}
\newglossaryentry{self-control}
{
  name={self-control},
  description={In general usage, restraint exercised over one's own impulses, emotions, or desires. In finance, referring to the ability to commit to future courses of action. Weak self-control implies that an individual is impulsive, incapable of honoring self-imposed rules, e.g. following a diet. Strong self-control implies that an individual is able to make rules and to follow them}
}
\newglossaryentry{semelparous}
{
  name={semelparous},
  description={A mode of reproduction in which parents die when offspring are born}
}
\newglossaryentry{semiparametric.model}
{
  name={semiparametric model},
  description={In statistics, a model that has parametric and nonparametric (infinite dimensional parameter space) parts}
}
\newglossaryentry{sentiment}
{
  name={sentiment},
  description={A term that refers to the range of opinions and attitudes of the players in a market. In financial markets, diverse attitudes are often simplified to: bullish, bearish and neutral; in real estate, to prices ``too high'', ``too low'' or ``afforable.'' Sentiment has been believed to be important in selecting market tops and bottoms --- low bullish sentiment predicting an advancing market, and high bullish sentiment predicting a declining market. Same as \gls{market.sentiment}}
}
\newglossaryentry{sentiment.indicator}
{
  name={sentiment indicator},
  description={A statistic that purports to measure bullish and bearish \gls{sentiment} in a market}
}
\newglossaryentry{SEO}
{
  name={SEO},
  description={An acronym for \emph{Seasoned Equity Offering}, which offers shares of stock denominated in dollars on an existing foreign publicly-held company. See also \gls{IPO}}
}
\newglossaryentry{shannon}
{
  name={Claude Shannon},
  description={A scientist who developed the theory of communication and information theory}
}
\newglossaryentry{shares.outstanding}
{
  name={shares outstanding},
  description={Same as \gls{outstanding.shares}}
}
\newglossaryentry{Sharpe.ratio}
{
  name={Sharpe ratio},
  description={The \emph{Sharpe ratio} for return series $R_t$, $t = 1, 2, \ldots, n$ equals the difference of its mean and the risk free rate divided by its standard deviation. The \gls{information.ratio} is similar but does not deduct the \gls{risk-free.rate} from the mean return}
}
\newglossaryentry{Shefrin.Hersh}
{
  name={Hersh Shefrin},
  description={Author of seminal research in behavioral finance explaining, inter alia, the disposition effect and investors' preferences for dividends}
}
\newglossaryentry{Shell}
{
  name={Shell},
  description={A petroleum company headquarted in Amsterdam, but which shares all profits with the English company, \gls{Royal.Dutch.Company}. Deviation of the prices of these two companies is the source of a serious EMH \gls{anomaly}}
}
\newglossaryentry{signal}
{
  name={signal},
  description={In the terminology used in this work, a \emph{signal} at time $t$ is a statistic calculated from data available on, or prior to $t$, for use in making trading decisions}
}
\newglossaryentry{signal.profile}
{
  name={signal profile},
  description={In the technical trading terminology ot this book, a \emph{signal.profile} or just \gls{profile} is a collection of vectors $v_t \in \mathbb{R}^n$, that at each time $t$ gives a trading signal for trading variables $p_t \in  \mathbb{R}^n$ as a scalar product $v_t \cdot p_t$. If $v_t = v$ is constant for all $t$, then the \emph{profile} is called \emph{stationary}. For the most part, methods used in this book produce stationary profiles}
}
\newglossaryentry{simon}
{
  name={Herbert Simon},
  description={Nobel Laureate in Economics, 1978}
}
\newglossaryentry{simple.moving.average}
{
  name={simple moving average},
  description={A simple moving average with memory $k$ is a \gls{moving.average} having $\alpha_s = 1/k$ for $0 \le s \le k-1$ and $0$ for $s ge k$, where $SMA_t = \sum_{s=0}^{s=\infty} \alpha_s X_{t-s}$}
}
\newglossaryentry{slippage}
{
  name={slippage},
  description={In the context of trade execution, the difference between the target price of a trade and the actual price of execution. For example, a purchase of $1,000$ shares at target price price $\$10.50$ might be executed instead at $\$10.60$, in which case the slippage is $\$0.10$. In general, the slippage will be greater the larger the trade, and this must be accounted for in the backtesting of any trading system}
}
\newglossaryentry{vernon.smith}
{
  name={Vernon Smith},
  description={Nobel Laureate in Economics, 2002 for his contributions to experimental econometrics. Smith conducted laboratory experiments to test the predictions of economic theories},
  sort={Smith}
}
\newglossaryentry{smoothing}
{
  name={smoothing},
  description={A method of producing from a training dataset $(x_i,y_i)$ sampled from a density satisfying a functional relationship $f(x) = E[ Y | X=x]$, an estimator $\hat{f(x)}$ that had controlled variation, e.g. that is, ``smoothness''}
}
\newglossaryentry{smoothing.spline}
{
  name={smoothing spline},
  description={A smoothing method for a spline that uses a penalty for non-smoothness (smoothness regularization) to produce an approximating function. It is useful when the data have noise or there are multiple y-values for each x-value}
}
\newglossaryentry{social.dilemma}
{
  name={social dilemma},
  description={A game in which all individuals gain if all cooperate, but any one of them gains by not cooperating when all the others do}
}
\newglossaryentry{sornette}
{
  name={Didier Sornette},
  description={Prominent geophysicist and econophysicist, author or coauthor on crash models (LPPL), power laws in economics and complex systems. Currently Chair of Entrepeneurial Risks at ETH, Switzerland}
}
\newglossaryentry{spa}
{
  name={SP/A},
  description={An acronym for \emph{Security-Potential/Aspiration}, a theory of choice under uncertainty proposed by Lola Lopes. Uncertainty leads to fear which is manifested as need for security. Lower perceived undertainty encourages hope which manifests as ``perceived potential.'' All decisions are constrained by aspirations}
}
\newglossaryentry{speculative.pump.and.dump}
{
  name={speculative pump-and-dump},
  description={A rational trading strategy that consists of buying (selling) in concert with other like-minded traders in order to create a self-sustaining trend that can later be liquidated profitably to uninformed traders. This strategy relies on some traders (the more, the better) who follow trends without recourse to fundamental information}
}
\newglossaryentry{standardization}
{
  name={standardization},
  description={A term that applies to a sample $X_1, X_2, \ldots, X_n$. The $X_i$ are said to be standardized (to $X_i'$) by $X_i' = \frac{X_i - \, \bar{X}}{S}$, where $\bar{X} = n^{-1} \sum X_i$ and $S = (n-1)^{-1} \sum (X_i - \bar{X})^2$ }
}
\newglossaryentry{stanley}
{
  name={H. Eugene Stanley},
  description={Prominent physicist (statistical mechanics) and indisciplinary scientist at Boston University}
}
\newglossaryentry{stationary}
{
  name={stationary},
  description={A property of stochastic processes that requires the finite dimensional distributions to be invariant under time shifts}
}
\newglossaryentry{statistical.arbitrage}
{
  name={statistical arbitrage},
  description={An non-pure \gls{arbitrage} having a positive expected return, e.g. index \gls{arbitrage}, \glslink{pairs.trading.strategy}{pairs trading}, options delta-neutral hedging}
}
\newglossaryentry{statistical.independence}
{
  name={statistical independence},
  description={A reference to a set of random variables that are \gls{independent}. Same as \gls{independence}}
}
\newglossaryentry{statistical.regularity}
{
  name={statistical regularity},
  description={A recurring pattern in the market data for tradable assets},
  plural={statistical regularities}
}
\newglossaryentry{Statman.Meir}
{
  name={Meir Statman},
  description={Author of seminal research in behavioral finance explaining, inter alia, the disposition effect and investors' preferences for dividends}
}
\newglossaryentry{status.quo.bias}
{
  name={status quo bias},
  description={A bias in which a person values the status quo (no change) more highly than superior alternatives},
  plural={status quo biases}
}
\newglossaryentry{stochastic.process}
{
  name={stochastic process},
  description={A collection $\{X_t\}$ indexed by times $t \in T$. The time domain $T$ can be discrete or continuous},
  plural={stochastic processes}
}
\newglossaryentry{stop.order}
{
  name={stop order},
  description={A market order to buy (sell) at a \emph{stop price} or higher (lower). If the order is not executed at the stop price, it becomes a \gls{market.order}. See also \gls{market.order} and {limit.order}}
}
\newglossaryentry{Strategic.Analysis.of.Markets.Method}
{
  name={Strategic Analysis of Markets Method},
  description={The \emph{\acrshort{aSAMM}} is a framework for developing trading systems by using game theoretic, strategic and statistical analysis. As such, the SAMM is one degree removed from flesh-and-blood humans, but in its favor, is amenable to game theoretic analysis}
}
\newglossaryentry{strategic.ecology}
{
  name={strategic ecology},
  description={A strategic ecology at a moment in time is a collection of trading strategies existant at that time},
  plural={strategic ecologies}
}
\newglossaryentry{strategic.evolution}
{
  name={strategic evolution},
  description={A process in which strategies change, often in a patterned way, over time. The Minsky-Kindleberger Model is one example of strategic evolution}
}
\newglossaryentry{strategic.form}
{
  name={strategic form},
  description={An \gls{artificial.mathematical.game} in which the \glspl{strategy} available to the players are elements of strategy sets $C_i$. These sets in simple cases are finite, but in many games are infinite, and may represent a complete specification of every choice a player might make in some game contingency. Also called the \gls{normal.form}}
}
\newglossaryentry{strategic.plan}
{
  name={strategic plan},
  description={An incomplete trading scheme or idea that has parameters which if specified convert it to a \gls{trading.algorithm}. A plan such as ``Buy in a bull market, sell in a bear market by ,'' is not a strategic plan as it stands, but could be one if parameters type=\{bull, bear, neither\}, asset=\{set of tradable assets\}, execution=\{open,close\}, and any other variables that are required to decide what and when to buy and sell, are added to its statement}
}
\newglossaryentry{strategic.plan.capacity}
{
  name={strategic plan capacity},
  description={A return schedule for the aggregate return of strategies that implement a \gls{strategic.plan}. The idea is that while a \gls{strategic.plan} admits polymorphic implementations, the commonality of parameters and the commmonality of estimation methods will cause all to be exposed to common factors}
}
\newglossaryentry{strategic.pump.and.dump}
{
  name={strategic pump-and-dump},
  description={A rational trading strategy that consists of buying (selling) from a fundamentalist-initiated trend in concert with other like-minded traders in order to create a self-sustaining trend that can later be liquidated profitably to uninformed traders. This strategy relies on some traders (the more, the better) who follow trends without recourse to fundamental information}
}
\newglossaryentry{strategy}
{
  name={strategy},
  description={(a) In \gls{game.theory}, a complete specification of the choices that a player would make under every possible game contingency, (b) In trading, a set of guidelines, a set of rules or an algorithm that a trader, investor or computer uses to buy and sell. Some strategies are purely mechanical and can be executed by a computer. Some are purely discretionary, and require one to treat human decisions to buy and sell as an algorithm, albeit one that cannot be programmed},
 plural={strategies}
}
\newglossaryentry{strategy.capacity}
{
  name={strategy capacity},
  description={A schedule that indicates a strategy's return degredation as a function of the aggregate position size}
}
\newglossaryentry{strategy.diversity}
{
  name={strategy diversity},
  description={In the \gls{Pursuit.of.Profits.Paradigm}, the degree of heterogeneity among strategies that implement its \gls{strategic.plan}}
}
\newglossaryentry{strategy.linkage}
{
  name={strategy linkage},
  description={In the \gls{Pursuit.of.Profits.Paradigm}, the degree of potentially contagious connection to other strategies}
}
\newglossaryentry{strategy.profile}
{
  name={strategy profile},
  description={In \gls{game.theory}, given strategy sets $C_i$ for $i =1, 2, \ldots, n$, a vector of strategy choices $c = (c_1, c_2, \ldots, c_n)$}
}
\newglossaryentry{strict.ordering}
{
  name={strict ordering},
  description={In utility theory, a strict ordering $\succ$ is derived from a weak ordering $\succeq$ by  
\begin{equation*}
  A \succ B \mbox{ } := \mbox{ } A \succeq B \mbox{ and } A \not\sim B
\end{equation*}
}
}
\newglossaryentry{t.distribution}
{
  name={Student's t distribution},
  description={Random variable $X$ has a Student's t distribution with $\nu$ degrees of freedom if its density is 
\[
  \frac{\Gamma(\frac{\nu+1}{2})}{\sqrt{\nu \pi} \, \Gamma(\frac{\nu}{2})} \, \left[ 1 + \frac{x^2}{\nu} \right]^{-\frac{\nu + 1}{2}} \text{ for } x \in (-\infty,\infty)
\]}
}
\newglossaryentry{stylized.fact}
{
  name={stylized fact},
  description={A \gls{statistical.regularity}}
}
\newglossaryentry{subadditivity.bias}
{
  name={subadditivity bias},
  description={The tendency to value a whole less than the sum of its parts}
}
\newglossaryentry{subdominant.paradigm}
{
  name={subdominant paradigm},
  description={A \gls{market.paradigm} which is a competitor to the \gls{dominant.paradigm}},
}
\newglossaryentry{subexponential}
{
  name={subexponential},
  description={A distribution that satisfies the asymptotic tail condition: 
  \[
    P \left[ (X_1 + X_2 + \cdots + X_n) > x \right] \sim n P\left[ X_1 > x \right]
  \]
  }
}
\newglossaryentry{submartingale}
{
  name={submartingale},
  description={In this work, a time series $X_t$ for which $E[X_t] < \infty$ and $E[X_{t+1} \, | \, X_t ] \ge X_t$ for all meaningful $t$. A series for which the inequality is strict is called a \emph{strict submartingale}}
}
\newglossaryentry{Substitution}
{
  name={Substitution},
  description={An axiom of \gls{cardinal.utility} theory which requires that for $A \sim B$ and any event $C$, the ithe lotteries $(p,A; (1-p)C)$ and $(p,B; (1-p)C)$ are equivalent for any $p \in [0,1]$, $(p,A; (1-p)C) \sim (p,B; (1-p)C)$}
}
\newglossaryentry{SUE}
{
  name={SUE},
  description={An acronym for \emph{Standardized Unexpected Earnings}, defined as the 
             current earnings minus those of a year ago (YoY earnings) divided by 
             their standard deviation}
}
\newglossaryentry{superior.active.portfolio}
{
  name={superior active portfolio},
  description={An active portfolio $A$ for which $E[A] > 0$}
}
\newglossaryentry{supermartingale}
{
  name={supermartingale},
  description={In this work, a time series $X_t$ for which $E[X_t] < \infty$ and $E[X_{t+1} \, | \, X_t ] \le X_t$ for all meaningful $t$. A series for which the inequality is strict is called a \emph{strict supermartingale}}
}
\newglossaryentry{support}
{
  name={support},
  description={Same as \gls{price.support}. See also \gls{price.resistance} and \gls{resistance}}
}
\newglossaryentry{suppressed.ambiguity}
{
  name={suppressed ambiguity},
  description={A mental process in which facts viewed more favorably, suppress facts that are in conflict with, or contradictory to them. The \gls{halo.effect} can lead to suppressed ambiguity}
}
\newglossaryentry{surprise}
{
  name={surprise},
  description={In the classification of market events, an unscheduled, unanticipated event that has market-moving impact. See also \gls{suspense}}
}
\newglossaryentry{survival.function}
{
  name={survival function},
  description={The function $S(x) = F_{\!_{>}}(x) = 1 - F_{\!_{\le}}(x)$, where $F_{\!_{\le}}(x)$ is a \gls{cumulative.distribution.function}. Also called the \gls{counter.cumulative.distribution.function}}
}
\newglossaryentry{suspense}
{
  name={suspense},
  description={In the classification of market events, one for which the time and place of occurrence is known, even inexactly, but not the exact outcome. In general, {suspense} events can cause short-term moves in a market due to the fact that most traders withdraw their orders before the event. Resting orders on the ``wrong'' side of the market can be ``picked off'' after the event. See also \gls{surprise}}
}
\newglossaryentry{System1}
{
  name={System 1},
  description={The mind's ``reflexive'' cognitive subsystem; massively parallel, quick, hard to reprogram, effortless}
}
\newglossaryentry{System2}
{
  name={System 2},
  description={The mind's ``reflective'' cognitive subsystem; single focussed, slow, relatively easy to reprogram, effortful}
}

\newglossaryentry{technical.analysis}
{
  name={technical analysis},
  description={The study of predictive patterns in financial markets involving only time, open, high, low, close, volume and open interest},
  plural={technical analyses}
}
\newglossaryentry{tell}
{
  name={tell},
  description={A giveaway mannerism, gesture or expression that often portends a rather unfortunate futurei, emphatically not related to \emph{fortune telling}}
}
\newglossaryentry{temporal.construal}
{
  name={temporal construal},
  description={The tendency to represent events near in time more concretely and in greater detail than events further removed}
}
\newglossaryentry{Thaler.Richard}
{
  name={Richard Thaler},
  description={Colleague of Daniel Kahneman and descoverer of behavioral finance terms --- loss aversion, the status quo bias and mental accounting}
}
\newglossaryentry{tick}
{
  name={tick},
  description={A unit of change in a futures contract. In Eurodollars priced to two decimal places, a tick is a change of $0.01$, e.g. $94.32$ to $94.33$. Generally, a tick is a minimum change, but that rule has many exceptions. Some Eurodollars, for example, are prices in \emph{half-ticks} ($0.005$) or \emph{quarter-ticks} ($0.0025$)}
}
\newglossaryentry{timestamped}
{
  name={timestamped},
  description={Referring to an event that is recorded as having occurred at a particular time, the ``timestamped'' time, of course}
}
\newglossaryentry{time.series}
{
  name={time series},
  description={A probabilistic process that produces observations at deterministic or random times}
}
\newglossaryentry{tower.property}
{
  name={tower property},
  description={Information sets $\{I_t\}$ have the tower property if $I_{t+1} \supset I_t$ for all meaningful $t$}
}
\newglossaryentry{tracking.error}
{
  name={tracking error},
  description={The deviation of the value of a \gls{basket.of.stocks} from a benchmark index}
}
\newglossaryentry{trade}
{
  name={trade},
  description={In this work, a \gls{founding.transaction} followed (eventually) by a \gls{closing.transaction} that exits the entire position created by the founding and any subsequent transactions that maintain a position on the same side of the market. This concept deviates slightly from trader's parlance, in that there may be many reductions in the size of the opening transaction before its complete closing. When refering to historical closed trades, is called a \emph{closed trade}, while when ongoing (not yet closed), is called an \emph{open trade}}
}
\newglossaryentry{Trader/Manager}
{
  name={Trader/Manager},
  description={A nonstandard term used in this work for a firm that invests client funds directly into tradable assets such as bonds, stocks and commodities. See also \gls{Manager.of.Managers}}
}
\newglossaryentry{trading.algorithm}
{
  name={trading algorithm},
  description={In trading and investment, a collection of rules for trading that are specific enough to be implemented as a computer program, requiring no human intervention beyond decisions of when to use them}
}
\newglossaryentry{trading.position}
{
  name={trading position},
  description={In a trading context, either a \emph{long position} in an instrument consists of ownership of the instrument or fungible contract that can be traded, or a \emph{short position} that consists of the opposite of a long position (except for financing costs). In stocks, a short position is  created synthetically through stock borrowing; in instruments that are two-sided tontracts, the opposite side of a long contract position. When the context is clear, abbreviated to \gls{position}. For a trading entity, its position is the collection of all positions in instruments}
}
\newglossaryentry{trading.system}
{
  name={trading system},
  description={In mechanical trading and investment, a collection of well-defined rules that specify when to enter the market, the size and side (buy or sell) for entry, and when to exit all or part of an existing position. In this work, the same as a \gls{mechanical.trading.system}}
}
\newglossaryentry{Trading.System.Summary}
{
  name={Trading System Summary},
  description={A statistical report that summarizes the historical activity of a trading system in order to evaluate its performance},
  plural={Trading System Summaries}
}
\newglossaryentry{trading.time}
{
  name={trading time},
  description={Market parlance for time elapsed only during periods in which trading is possible. The daytime trading session at the \acrshort{aNYSE}, for example, is from $9$:$30$ to $4$:$30$ EST, Mondays through Fridays, excepting designated holidays and market closings due to extraordinary events}
}
\newglossaryentry{tragedy.of.the.commons}
{
  name={tragedy of the commons},
  description={A game paradigm in which many players of a game compete for a free or semi-free good in scarce supply, resulting in its uncoordinated depletion as each player attempts to maximize profit independently of other players}
}
\newglossaryentry{transaction}
{
  name={transaction},
  description={A \gls{transaction} is an executed or partially executed order, that is, one in which a quantity of a financial instrument bought or sold at a particular time and place}
}
\newglossaryentry{transactional.data}
{
  name={transactional data},
  description={Observational data of a \gls{transaction} event, in this work a buy or sell, that occurs at a particular time and is timestamped. Other market events, such as the posting of a buy or sell order or cancellation of same is not transactional data, although it becomes so if and when an actual purchase or sale is realized}
}
\newglossaryentry{transactional.format}
{
  name={transactional format},
  description={The representation of a trading history as a series of time-stamped transactions}
}
\newglossaryentry{Transitivity}
{
  name={Transitivity},
  description={An axiom of utility theory that requires of events $A \succ B$ and $B \succ C$, that $A \succ C$}
}
\newglossaryentry{trend}
{
  name={trend},
  description={In \gls{technical.analysis}, a venerated \gls{statistical.regularity} in which many portfolios that have appreciated have the tendency to continue appreciating, and many that have depreciated have a tendency to continue depreciating}
}
\newglossaryentry{trend.follower}
{
  name={trend follower},
  description={A trader that follows trends, that is, who buys when price is going up and/or sells when price is going down}
}
\newglossaryentry{trend.following}
{
  name={trend following},
  description={A strategy premised on the belief that profits are possible by detecting trends in progress and following them}
}
\newglossaryentry{TRIN}
{
  name={TRIN},
  description={The TRader's INdex (or ARMS index), which is calculated as the number of advancing issues divided by the number of declining issues}
}
\newglossaryentry{turnover.ratio}
{
  name={turnover ratio},
  description={For stocks, the ratio of the quantity traded over a specified period to the shares outstanding. For example, the daily turnover ratio of stock XYZ is the daily volume of XYZ divided by its shares outstanding}
}
\newglossaryentry{tversky}
{
  name={Amos Tversky},
  description={Amos Tversky, co-investigator with 2002 Nobel Laureate Daniel Kahneman. The ``T'' in the acronym ``K\&T''}
}

\newglossaryentry{unanticipated.news}
{
  name={unanticipated news},
  description={In financial markets, a piece of news, some portion of which was not anticipated. The term is often used to explain why some news stories have little market impact, while other seemingly innocuous stories have great impact. In this view, the market discounts all the ``anticipatible'' parts of news, and is affected only by the ``unanticipatible'' part}
}
\newglossaryentry{uncertain.decision.problem}
{
  name={uncertain decision problem},
  description={A \gls{decision.problem} that has some events with unknown probabilities. See also \gls{risky.decision.problem}}
}
\newglossaryentry{uncertainty}
{
  name={uncertainty},
  description={In financial decision theory, referring to situations affected by events whose probabilities are unknown. See also \gls{risk}}
}
\newglossaryentry{uncorrelated.white.noise}
{
  name={uncorrelated white noise},
  description={Same as \gls{white.noise}}
}
\newglossaryentry{underreaction}
{
  name={underreaction},
  description={In financial markets, underreaction refers to an insufficient price move in response to an event. See also \gls{overreaction}}
}
\newglossaryentry{upsizing.transaction}
{
  name={upsizing transaction},
  description={In a \gls{PPGS.renewal.process}, a \gls{transaction} that increases the absolute value of a current, nonzero \gls{position}}
}
\newglossaryentry{utiles}
{
  name={utiles},
  description={The units of a \gls{utility.function}}
}
\newglossaryentry{utility.function}
{
  name={utility function},
  description={A function $U(A)$ that expresses the six axioms of \Gls{cardinal.utility}. Such a function is unique up to a positive affine transformation. The expected values of such a function are the basis of \gls{vNM.expected.utility.theory}. See \gls{utility.theory}}
}
\newglossaryentry{utility.theory}
{
  name={utility theory},
  description={A theory of consistent choice making which has two types: (1) ordinal, and (2) cardinal. Ordinal utility theory has axioms of \gls{Complete.Ordering}, \gls{Reflexivity.Of.Preference} and \gls{Transitivity}. Cardinal utility theory requires the ordinal axioms and those of \gls{Compound.Equivalence}, \gls{Substitution} and the \gls{Continuity.Axiom}}
}

\newglossaryentry{valuation}
{
  name={valuation},
  description={In general, the process of assigning a value, e.g. in finance, valuation of a financial instrument, a portfolio, an outcome, etc. In \gls{prospect.theory}, the process of valuing a \glslink{framing}{framed} prospect using a \gls{value.function} and \gls{weight.function}}
}
\newglossaryentry{value}
{
  name={value},
  description={In game theory, the average payoffs that players achieve with best play. \acrshort{avnm} proved that every \gls{zero.sum} randomized game has a value, although some non-zero sum games have no well-defined one}
}
\newglossaryentry{value.function}
{
  name={value function},
  description={In \gls{prospect.theory}, a function that describes value of a decision relative to a reference point, which \acrshort{aKnT} demonstrated empirically to be convex below the reference point and concave above it}
}
\newglossaryentry{value.investing}
{
  name={value investing},
  description={An method of security analysis that uses market fundamentals to find \emph{investment value}. Made famous by Benjamin Graham in his landmark book, ``The Intelligent Investor'' \cite{graham2003intelligent}}
}
\newglossaryentry{verification.stage}
{
  name={verification stage},
  description={The stage of scientific inquiry that confirms hypotheses by collecting new data and showing that predictions match theory}
}
\newglossaryentry{vNM.expected.utility.theory}
{
  name={von Neumann/Morgenstern Expected Utility Theory},
  description={The axiomatic expected utility theory of John von Neumann and Oskar Morgenstern, which assumes that decision making is preference-consistent, extensible to lotteries and Archimedian}
}
\newglossaryentry{vNM.rational}
{
  name={vNM rational},
  description={An actor is vNM rational if he, she or it uses \gls{vNM.expected.utility.theory} to make decisions. In our usage, a vNM rational actor is simply called ``rational''}
}
\newglossaryentry{volatility.clustering}
{
  name={volatility clustering},
  description={Referring to episodic low and high volatility in financial time series in financial time series. Same as volatility \gls{intermittency}}
}

\newglossaryentry{weak.ordering}
{
  name={weak ordering},
  description={In utility theory, a relation that satisfies the three axioms: \gls{Complete.Ordering}, \gls{Reflexivity.Of.Preference} and \gls{Transitivity}. In this work, called a \gls{preference.order}}
}
\newglossaryentry{weight.function}
{
  name={weight function},
  description={In \gls{prospect.theory}, a function that maps probabilities in a prospect to weights useful in \gls{valuation}}
}
\newglossaryentry{weighted.moving.average}
{
  name={weighted moving average},
  description={A weighted moving average with memory $k$ is a \gls{moving.average} having $\alpha_s \ge 0$ for $0 \le s \le k-1$, and $\alpha = 0$ for $s ge k$, $\alpha_{k-1} > 0$ and $\sum_{s \ge 0} \alpha_s = 1$, where $WMA_t = \sum_{s=0}^{s=\infty} \alpha_s X_{t-s}$}
}
\newglossaryentry{white.noise}
{
  name={white noise},
  description={A time series $X_t$ for which $E[X_t] = 0$,
               $Var[X_t] = \sigma^2$, and $Cov[X_s,X_t] = 0$ 
               for all meaningful $s, t$ and where $\sigma$ 
               is the standard deviation of $X_1$}
}
\newglossaryentry{winner's.curse}
{
  name={winner's curse},
  description={A game-theoretic phenomenon in competitive auction markets. In effect, it states that the winner of a multi-bidder auction will be the one that is most optimistic, thus will overpay. In game theory, analogs of the winner's curse occur in all \emph{coordination games}}
}
\newglossaryentry{wn}
{
  name={WN},
  description={Abbreviation for white noise. Sometimes 
               written as $WN(\mu,\sigma^2)$, which enlarges 
               the definition to include processes that have 
               non-zero drift $\mu$}
}
\newglossaryentry{worst.loss.function}
{
  name={worst loss function},
  description={For a real-valued time series $\{X_s\}_{s=1}^{s=t}$ of gains, the largest cumulative loss $\mathcal{L}(X,t)$ to time $t$, i.e. for $V_0 = 0$, $V_s = \sum_{u=1}^{u=s} X_u$, $s \ge 1$, \[\mathcal{L}(X,t) = -\underset{1 \le s \le t}{min} V_s.\] Note that $\mathcal{L}(X,t) \ge 0$}
}

\newglossaryentry{zero.sum}
{
  name={zero-sum},
  description={A real or artificial mathematical game in which the sum of payoffs to the players is zero}
}
\newglossaryentry{Zipfs.law}
{
  name={Zipf's law},
  description={A power law distribution having positive lower threshold and exponent $1$}
}

\newacronym{aaa}{$A\&A$}{Anchoring \& Adjustment}
\newacronym{aACE}{ACE}{Alternating Conditional Expectations Model}
\newacronym{aAMEX}{AMEX}{American Stock Exchange}
\newacronym{aarch}{$ARCH$}{Autoregressive Conditional Heteroschedasticity}
\newacronym{aarima}{$ARIMA$}{Autoregressive Integrated Moving Average}
\newacronym{aarma}{$ARMA$}{Autoregressive Moving Average}
\newacronym{aAMH}{AMH}{Adaptive Markets Hypothesis}
\newacronym{aAVAS}{AVAS}{Additivity and  Variance Stabilizing Transformations Model}
\newacronym{abf}{BF}{behavioral finance}
\newacronym{abp}{b.p.}{basis point}
\newacronym{abnl}{B\&L}{Brennan \& Lo}
\newacronym{aCAPM}{CAPM}{Capital Asset Pricing Model}
\newacronym{aCAR}{CAR}{Cumulative abnormal return}
\newacronym{aCARA}{CARA}{Constant Absolute Risk Aversion}
\newacronym{accdf}{CCDF}{counter-cumulative distribution function}
\newacronym{acdf}{CDF}{cumulative distribution function}
\newacronym{aCME}{CME}{Chicago Mercantile Exchange}
\newacronym{aCRRA}{CRRA}{Constant Relative Risk Aversion}
\newacronym{acrsp}{CRSP}{Center for Research in Security Prices}
\newacronym{aCPT}{CPT}{Cumulative Prospect Theory}
\newacronym{aDJIA}{DJIA}{Dow Jones Industrial Average}
\newacronym{adp}{DP}{Decision Problem}
\newacronym[longplural={Decision Theories}]{adt}{DT}{Decision Theory}
\newacronym{aefm}{EFM}{Evolutionary Finance Model}
\newacronym{aEMH}{EMH}{Efficient Market Hypothesis}
\newacronym{aETF}{ETF}{Exchange Traded Fund}
\newacronym{aeu}{EU}{Expected Utility}
\newacronym{afd}{FD}{Fast-decaying Distribution}
\newacronym{aFED}{FED}{Federal Reserve System}
\newacronym{aFIH}{FIH}{Financial Instability Hypothesis}
\newacronym{aFLOG}{FLOG}{Fundamental Laws of Gambling}
\newacronym{afnd}{F\&D}{Froot \& Debora}
\newacronym{afifo}{FIFO}{First In/First Out}
\newacronym{aGAM}{GAM}{Generalized Additive Model}
\newacronym{aGFC}{GFC}{Global Financial Crisis of 2007-8}
\newacronym{aGnH}{G\&H}{Grinblatt \& Han}
\newacronym{aGnT}{G\&T}{Gigerenzer \& Todd}
\newacronym{agarch}{$GARCH$}{Generalized Autoregressive Conditional Heteroschedasticity}
\newacronym{ahbm}{HBM}{Heterogeneous Beliefs Model}
\newacronym{aHE}{HE}{Homo Economicus}
\newacronym[longplural={independent components analyses}]{aICA}{ICA}{independent components analysis}
\newacronym{aIRA}{IRA}{Individual Retirement Account}
\newacronym{aJnT}{J\&T}{Jegadeesh \& Titman}
\newacronym{alor}{LOR}{Leland O’Brien Rubinstein Associates}
\newacronym{ahft}{HFT}{High-frequency Trading}
\newacronym{aKnT}{K\&T}{Kahneman and Tversky}
\newacronym{aMBS}{MBS}{Mortgage-Backed Security}
\newacronym{amkm}{MKM}{Minsky-Kindleberger Model}
\newacronym{aMNS}{MNS}{Market Neurtal Strategy}
\newacronym{aNASDAQ}{NASDAQ}{National Association of Security Dealers Exchange}
\newacronym{ancm}{NCM}{Noisy Channel Model}
\newacronym{amna}{M\&A}{Mergers \& Acquisitions}
\newacronym{aNYSE}{NYSE}{New York Stock Exchange}
\newacronym[longplural={principal components analyses}]{aPCA}{PCA}{principal components analysis}
\newacronym{apd}{PD}{Prisoner's Dilemma}
\newacronym{apdf}{p.d.f}{probability density function}
\newacronym{apgg}{PGG}{Public Goods Game}
\newacronym{apl}{PL}{Power Law}
\newacronym{aPLT}{PLT}{Power Law Tail}
\newacronym{aPPGS}{PPGS}{Potentially Profitable Gambling System}
\newacronym{aptb}{PTB}{Pick-the-Best Heuristic}
\newacronym{aRE}{RE}{Rational Expectations}
\newacronym{arecap}{RECAP}{Record, Evaluate, and Compare Alternative Prices}
\newacronym[longplural={Rational Belief Equilibria}]{arbe}{RBE}{Rational Belief Equilibrium}
\newacronym{arct}{RCT}{Rational Choice Theory}
\newacronym{armm}{RMM}{Reflexive Market Model}
\newacronym{aSAMM}{SAMM}{Strategic Analysis of Markets Method}
\newacronym{asf}{s.f.}{survival function}
\newacronym{aSnP}{S\&P 500}{Standard \& Poors 500 Index}
\newacronym{aSMK}{SMK}{Soros-Minsky-Kindleberger Model}
\newacronym{aAUM}{AUM}{Assets Under Management}
\newacronym{aVIX}{VIX}{CBOE Volatility Index}
\newacronym{avnm}{vNM}{von Neumann/Morgenstern}
\newacronym{aweird}{WEIRD}{Western, Educated, Industrial, Rich and Democratic}
\newacronym{awn}{WN}{white noise}
\newacronym{aWSMC87}{Crash of 1987}{Worldwide Stock Market Crash of 1987}
\newacronym{awysiati}{WYSIATI}{What you see is all there is}

\begin{document}

% Add the title section.
\begin{titlepage}
  \centering
  {\scshape\Large {On a Constructive Theory of Markets \\ \vspace{5mm} \normalsize Summarized from Volume 1 of \\``The Strategic Analysis of Financial Markets''}\par}
  \vspace{1.5cm}
  {\scshape\large April 1, 2017\par}
  \vspace{2cm}
  {\large Steven D. Moffitt\textsuperscript{\textdagger}\par}
  \vfill
  \textdagger Adjunct Professor of Finance, Stuart School of Business, Illinois Institute of Technology and
              Principal, Market Pattern Research, Inc.\par
\end{titlepage}

% Add an abstract.
\abstract{
\noindent 
This article is a prologue to the article ``Why Markets are Inefficient: A Gambling 'Theory' of Financial Markets for Practitioners and Theorists'' (\cite{moffitt:SSRN:WhyMarketsAreInefficient:2017}). It presents important background for that article --- why gambling is important, even necessary, for real-world traders, the reason for the superiority of the strategic/gambling approach to the competing market ideologies of market fundamentalism and the scientific approach, and its potential to uncover profitable trading systems. Much of this article was drawn from Chapter 1 of \cite{moffitt2017V1}.
}

\null\vfill
\newpage

\begin{epigraphs}
\qitem{In theory, there's no difference between theory and practice. 
          In practice, there is.}%
       {Walter J. Savitch, computer scientist and author of \emph{Pascal: An Introduction to the Art and Science of Programming (1984)}. Often incorrectly attributed to Johannes Snepsheut or Yogi Berra.\footnote{Thanks to Harry Markowitz for correcting the original attribution.}}
\qitem{
      Economics consists of theoretical laws which nobody has verified and of empirical
      laws which nobody can explain.
      }%
      {epigram of Michal Kalecki, quoted by \cite{steindl1965random}, p18.}

\end{epigraphs}

\section{A Practical Approach}\label{OACTOM:S:APracticalApproach}

Behind most models of mathematical finance lie some really astounding assumptions \ldots that everybody is rational, that the law of one price holds, that no persistent \gls{arbitrage}\index{arbitrage} is possible, and so on. And most introductions to investing are either revelations of secret methods for making millions, or recitations of the immutable laws of cash flows, \glspl{risk-adjusted.return}\index{risk-adjusted return}\index{return!risk-adjusted} and efficient portfolios. But little of this literature offers credible, general principles to an investor trying to get an edge in the investment game\index{games \& game theory}. Get rich schemes fail the sniff test --- why are these precious secrets being peddled for \$34.95? And most ``investing theory'' can be trusted only if its assumptions align with reality (which they do not). The received wisdom of finance today, to paraphrase something I was told ages ago, is ``a magic show and a bag of tricks.''
 
Mainstream economic theory's markets are populated by rational automatons, but real markets are populated by human beings. Emotions are important. Mistakes abound. And indeed, markets are no more efficient than the emotional, mistake-ridden people that comprise them. To emphasize this point, consider a little tale of the economist and the gambler.

\section{A Tale of the Economist and the Gambler}\label{OACTOM:S:TheEconomistAndTheGambler}

An economist and a gambler are presented a puzzle: ``A fair coin turns up heads on \(20\) consecutive flips. What's the probability of this outcome?" The economist, practiced in the art of probability, answers ``\(1\) in \(1,048,576\).'' But the gambler answers, quizzically, ``It's not a fair coin!'' 

On the surface of it, the economist's answer seems more sensible since he interprets the puzzle literally, imagining that fair coins, that is, ones having a probability of exactly \(1/2\), exist. He reasons that all sequences of \(20\) heads and tails are equally likely, implicitly assuming independent flips. Thus he arrives at the answer ``\(1\) in \(1,048,576\).'' This standard reasoning has been taught to millions of students. But how many realize its deficiencies?

First, fair coins do not exist in the real world. Ones that are effectively fair can be manufactured through contrivance and ingenuity, but they do not exist in nature nor among coins minted in the United States. Is an United States penny minted in 1960 fair? Answer: Indeterminate without additional information, but under reasonable assumptions explained below, no! \footnote{The Lincoln Head penny of $1960$ emphasizes that a real coin, not a hypothetical one, is unfair. Other years in the Lincoln series could have been specified interchangeably.} 

Second, the act of ``flipping'' is ignored in the economist's version, but in reality, that act is of crucial importance. If, for example, the coin is flipped with low angular momentum, then the result will be strongly biased toward the face that is upward when the coin is flipped! That is, if the upward face is heads, then heads will occur more often than tails, but if the upward face is tails, then tails will occur more often (\cite{citeulike:1269501})! And not only does the bias persist when larger angular momenta are imparted to the coin (the upward face lands up about \(51\%\) of the time with reasonable assumptions on an upper limit to angular momentum), but in theory and in practice it never dies out completely! On the other hand, if a U.S. penny minted in 1960 is spun on its side with a flick of the finger on a level, flat surface, the probability of tails is about \(0.70\), that of heads \(0.30\). Clearly, then, the instructions on how to randomize the coin, as well as initial conditions, will affect the outcome.

Here is a little story that actually happened in San Francisco in 1985. A friend of mine who was an options market maker and champion backgammon player was approached by a man who offered him offered him \(3\):\(2\) odds on flips of a coin. My friend would win, say, \(\$15\) when he called the coin correctly, but pay only \(\$10\) when he was wrong. He was what I'd term a ``mathematical gambler,''  accustomed to playing against people, both in the options pits and in backgammon, who made inferior ``moves.'' When he thought he had an edge, he was quick to pounce. So the ``Flipper'' began calling the coin that my friend flipped and my friend won a little at first. When the flipper then asked to escalate the bet, he agreed. After a bit of back and forth, the bet size was further escalated, at which point my friend promptly went on a losing streak that cost $\$15,000$! The Flipper was a ``confidence man,'' or ``con man'' for short, who gained victims' confidences in order to rip them off. He appeared for a while in San Francisco and stung a few traders there, and then vanished as mysteriously as he'd appeared. There is a saying in the markets and in life --- if it's too good to be true, it isn't. 

Unlike the economist, the gambler imagines ``If I had the opportunity to bet on head or tail of a coin that has just come up heads twenty consecutive times, what would I do?'' In this interpretation of the puzzle, few would disagree -- ``Bet heads on the next flip,'' which from a gambler's perspective, implies that the coin is unlikely to be fair.

This difference in approach can be cast in familiar terms. An economist is a probabilist, a gambler, a statistician. No people enter the financial theorist's thought; people are always present in the gambler's. In the ``20 heads of a fair coin,'' the gambler admits the possibility that the puzzle's presenter is a liar or misinformed, i.e. that the coin is in actuality unfair. 

\section{Why Speak of Gambling?}\label{OACTOM:S:WhySpeakOfGambling}

I feel obliged to explain why gambling is being discussed in a serious article about financial markets. A popular view holds that gambling is a vulgar activity that attracts the low-minded, thus having no redeeming social value. A friend even joked, ``Having gambling on your resume is a plus only if you want to be a trader.'' Gambling is often thought of as reckless betting, and this caricature is reinforced by movies of reckless gamblers who lead roller coaster lives, an anathema to prudent people.

Although ill-favored ``gambling'' and respectable ``investing'' have much in common, the ``sophisticated'' wealth creation community eschews any mention of gambling. Nor is the vulgar gambling idiom, ``bankroll,'' ever heard in the stead of ``wealth'' or ``betting'' for ``trading.'' And academic finance, by and large, has adopted the polite terms, as if they make the entire enterprise of investing more acceptable.

Consider this anecdote. My friend Harold, a successful gambler I'd known for many years was getting married. Several of his old gambling buddies, myself included, sat together at the ceremony as the Rabbi introduced the bride and groom. Julia, he said, was from a respectable family and was honorably employed as a teacher. But Harold's avocation, he said, was ``professional risk taking.'' The word ``respectable'' and ``gambler'' (Is ``respectable gambler'' an oxymoron?) did not pass his lips.

Such genteelisms are amusing to  \gls{professional.gamblers}\index{professional gamblers}\index{gambler, professional} who believe, correctly in my opinion, that  high-minded terms like ''investing'' and ''wealth creation'' are designed for ``suckers,'' those investors who na\"{i}vely believe that their money is in the hands of prudent managers. And this is a dangerous mirage indeed, since ``prudent'' investment managers are mostly wandering in the dark with penlights searching for winners, just like everybody else, and that as a group they achieve sub par returns.

There is another illusion at work, though. While there are wild gamblers who leave their fates to the gods, such is most assuredly not true of \glslink{professional.gamblers}{professional gamblers}\index{professional gamblers}\index{gambler, professional}. The principles followed by professional gamblers are as simple to state as they are hard to follow: (1) never bet without an \gls{edge}\index{edge}, and (2) bet an amount (possibly zero) appropriate for that edge.
%will be discussed in Section \ref{GG:S:GamblingPrinciplesAndGrationality} which will make it abundantly clear that gambling principles are based on sound reasoning, not on seat-of-the-pants betting.

The paradigm of ``\(20\) heads'' that opened this article is emblematic of the difference between the economist and the gambler, pointing to differences in motivation and methods. The economist seeks to understand how markets work, but has a bias that markets can't be beaten consistently. Since the economist accepts that markets can't be beaten, he conducts  no serious investigation of that possibility. The gambler on the other hand, also cares about understanding markets, but only for insight on how to gain an edge. And for the gambler, it is more important to infer that someone made a bad bet than to understand why.

This article is about the gambler, not the economist. And about unfair coins, the ``Flipper,'' and the general question of how to win.

\section{On ``Economic Arguments'', Efficiency and Equilibrium}\label{OACTOM:S:OnEconomicArguments,EfficiencyAndEquilibrium}

Much of modern financial economics rests on the assumption that most people are rational.\index{rationality} But what if that assumption is wrong? Standard equilibrium arguments of the form ``the price x of an asset must be y because if it were higher, (rational) traders would sell it down toward y, or if it were lower, (rational) traders would buy it up toward y'' are then invalid. If this principle fails, the \emph{\gls{no.arbitrage.principle}} is unjustifiable. The argument for \acrshort{aCAPM}\index{CAPM} likewise fails. Markets can still be efficient without \gls{rationality}\index{rationality}, but the usual justification that relies on rational arbitrageurs is unavailable. Financial economics makes such pervasive use of rationality and by extension, equilibrium arguments, that it's not clear how much of current theory survives without them. 

The basic problem with many ``economic'' arguments is that they argue by contradiction --- ``if such and such were not true, then market magic would cause it to become true.'' I reject this sort of thinking as \emph{magical}, not \emph{constructive}, thinking. It uses a mathematical form of argument by contradiction, without the mathematical justification of that form --- namely that a proposition can either be true or false without any exceptions. The form would not be a problem if the clause ``because if it were not true, then somebody would make it true'' holds, but there are obvious logical problems with such statements. For example, is it even possible for people to devise a remedy, i.e. is there path or strategy whereby the remedy would be manifested? This is simply an existence question --- does a remedy even exist? Even if a remedy exists, will someone bother to employ it? Moreover, is a phenomenon obvious enough to be recognized by somebody? How can ``new'' phenomena be discovered if they are already known? These and other\ troubling aspects of the magical approach lead me to reject it in favor of a constructive approach. 

Now back to the main thought: if rationality is absent, can analytic understanding of financial markets be attained only through the conjuring of statistical magic? 

I think not. \label{PredictableIrrationality} Markets can be understood only by dropping the assumptions of \gls{rationality}\index{rationality} and efficiency in their extreme forms, and showing that markets still have an inherent order and inherent logic. But that order results primarily from the `` predictable irrationality'' of investors, as well as from people's uncoordinated attempts to profit. The market patterns that result do not rely on rationality or efficiency. In particular, there is a general failure to understand that myopic, self-interested actions often lead to self-defeating outcomes, despite the fact that this has always been a basic feature of markets (\cite{doi:10.1080/1351847X.2011.601872}). It is amazing that this feature is so ill understood, that it is little discussed in a literature that purports to describe real markets. The companion article \cite{moffitt:SSRN:WhyMarketsAreInefficient:2017} offers a framework that captures the order and logic of markets, and which argues that \gls{rationality}\index{rationality} and efficiency are routinely violated.

\section{Approaches to Understanding Markets}\label{OACTOM:S:ApproachesToUnderstandingMarkets}

A shortcoming of the financial literature is its general neglect of the topic of ``market philosophies.'' Roughly speaking, a market philosophy offers a simple (and often quite inadequate) paradigm of how markets operate. Most market participants, whether consciously or not, subscribe to a market philosophy. Therefore, it is important to understand these philosophies, the thesis being that as viewports through which investors see the financial world, they are essential to understanding its operation. Three philosophies and one major variant thereof are discussed below.

\subsection{The Idealistic Approach and Market Fundamentalism}\label{OACTOM:SS:TheIdealisticApproach}\index{philosophical approaches!idealistic|textbf}\index{market fundamentalism|textbf}

Mainstream finance in much of the last half of the twentieth century was mainly an idealistic study.  It postulated that people are rational without systematically studying their actual behavior, then developed theory from that assumption --- an approach often called \gls{market.fundamentalism}. The efficient market hypothesis\index{efficient market hypothesis (EMH)} (\acrshort{aEMH}\index{efficient market hypothesis (EMH)}), for example, assumes that almost all people are rational and that they unanimously agree on prices, so that fair prices result. The development of the \acrshort{aEMH}\index{efficient market hypothesis (EMH)} was not based on direct observation of people acting rationally, but rather from studies of how prices evolve and how markets react to news. But there are two major problems with this approach. First, one cannot conclude that because the interactions of many people produces certain behavior in markets, e.g. martingale prices, that those behaviors resulted from \gls{rationality}\index{rationality}. There can be other individual behaviors that could produce similar market outcomes. Second, that markets act as if rational people produced the prices is quite difficult to test. %On this we shall say more in Chapter \ref{C:ReexaminingMarketEfficiency}.

Regardless of its theoretical value, the \acrshort{aEMH}\index{efficient market hypothesis (EMH)} did capture one aspect of agreement between economists and investors --- that the market is hard to beat. Unfortunately, economists expressed this idea in the most extreme fashion possible, claiming that the market is impossible to beat except by \glspl{arbitrageur}\index{arbitrage}\index{arbitrageur}. The fact that many famous non-arbitrageur investors and speculators have earned superior returns over their entire careers is not prima facie evidence against the \acrshort{aEMH}\index{efficient market hypothesis (EMH)}, because a few winners should be expected by chance alone.

However, as more evidence against the \acrshort{aEMH}\index{efficient market hypothesis (EMH)} accumulated in the $1990$'s, its defenders relied increasingly on the argument that hidden risk underlay anomalous phenomena. Thus the mainstay of the \acrshort{aEMH}\index{efficient market hypothesis (EMH)}, \acrshort{aCAPM}, guarantees the fairness not of returns, but of \glspl{risk-adjusted.return}\index{risk-adjusted return}\index{return!risk-adjusted}, so high returns are natural when risk is high. But of course, high losses are also be expected when risk is high --- in this view, anomalistic winners were just lucky.

But some anomalies are immune to this argument. \cite{RePEc:bla:jfinan:v:46:y:1991:i:1:p:75-109} showed that the \Gls{closed-end.discount.puzzle}\index{closed-end discount puzzle} could not be argued away by high risk, while a behavioral paradigm offers an intuitively satisfying explanation. Despite the argument that a few will be lucky, it is also difficult to reconcile the \acrshort{aEMH}\index{efficient market hypothesis (EMH)} with the consistent superior \glspl{risk-adjusted.return}\index{risk-adjusted return}\index{return!risk-adjusted} of famous investors such as Warren Buffet and speculators such as George Soros.

Facing increasing evidence against the \acrshort{aEMH}\index{efficient market hypothesis (EMH)}, two of its long-time supporters capitulated, saying in effect, ``The market's not as efficient as we thought.''  But that's a story to be told elsewhere. 

\subsection{The Scientific Approach}\label{OACTOM:SS:TheScientificApproach}\index{philosophical approaches!scientific|textbf}\index{scientific approach|textbf}

The first thorough scientific studies of financial theory were initiated by psychologists \gls{kahneman}\index{Kahneman, Daniel}, \Gls{tversky}\index{Tversky, Amos} (\acrshort{aKnT}) and \gls{vernon.smith}\index{Smith, Vernon}. Over a period of years, \acrshort{aKnT} performed numerous experiments to learn how people make financial decisions.  In $1979$, \acrshort{aKnT} produced a general formula for their \gls{prospect.theory}\index{prospect theory} based on the average behavior of experimental subjects. It showed convincingly that individuals' decisions are described by \gls{prospect.theory}, not \gls{utility.theory}. Nonetheless, their theory falls short of the ideal, for it fails to produce unambiguous predictions. But by showing that human behavior is hardly random, it diminishes the \emph{\gls{noise.trader}}\index{efficient market hypothesis (EMH)!noise trader model} pillar of efficient markets\index{efficient market hypothesis (EMH)}.

The most disheartening thing from a scientific standpoint is that \acrshort{aKnT}'s work was for many years ignored by economic theorists, and \gls{rational.expectations}\index{rational expectations}\index{rationality} (\acrshort{aRE}\index{rationality}\index{rational expectations}) theories continued to be expounded as if nothing had changed. But since the late $1980$'s things have been changing, due to the discovery of numerous efficient market anomalies. In addition, markets events have made it increasingly hard to justify market efficiency, e.g. the global market crash of $1987$\index{crashes!Global Crash of $1987$}, the \gls{Dot-com.Bubble}\index{crashes!.COM Crash of 2000}\index{bubbles!Dot-Com Bubble of 1996-2000} of the late $1990$'s and the \glslink{Global.Financial.Crisis}{Global Financial Crisis of 2007-8}\index{crashes!Global Financial Crisis of 2007-8}\index{crashes!Global Financial Crisis of 2007-8}. 

\gls{vernon.smith} pioneered the study of economic predictions using laboratory experiments, a field often called \gls{experimental.economics}\index{experimental economics}. In important research, Smith studied how participants in auctions respond to incentives and institutional structure. He found in early experiments that experimental equilibria were close to those predicted by theory, and that institutional structure affected the speed of convergence. But in the mid $1980's$, \cite{RePEc:ecm:emetrp:v:56:y:1988:i:5:p:1119-51} and colleagues developed more sophisticated protocols, discovering that bubbles and crashes occurred routinely! And those findings have been corroborated again and again. Nonetheless, in a tribute to the tenacity (if not wisdom) of true believers, those findings were discounted because they often used undergraduate college students, because they were simplified versions of real markets, and so on.

But not all economists before the late $1980$'s were so myopic as to ignore experimental findings.  \gls{Shefrin.Hersh}\index{Shefrin, Hersh}, \gls{Statman.Meir}\index{Statman, Meir} and \gls{Thaler.Richard}\index{Thaler, Richard}, among others, began exploring the implications of \gls{prospect.theory}\index{prospect theory} to understand investor behavior. They found that it explained common features of financial decision making --- Shefrin explained investors' preferences for dividends in equities and the \gls{disposition.effect}\index{heuristics \& biases!disposition effect}\index{Disposition Effect} and Thaler introduced \gls{mental.accounting}\index{heuristics \& biases!mental accounting}\index{Mental Accounting} and \glspl{status.quo.bias}\index{heuristics \& biases!status quo bias}\index{Status Quo Bias}. Today, behavioral finance is an accepted field of economics, but unfortunately, has yet to fully penetrate the culture and training of economists and financial professionals.

Physicists began serious studies of markets in the $1990$'s, calling their field ``econophysics.'' Their approach was quite different from economists'.  In general, scientists start by discovering ``empirical regularities'' of a system --- this is the \gls{observational.experimental.stage}\index{science!experimental stage} of the scientific method.  Scientists then develop theories to explain these regularities; the criterion for a valid theory is that it must make unambiguous predictions. This is the \gls{hypothesis-forming.stage}\index{science!hypothesis-forming stage} of the scientific method.  Each hypothesis so formed is then tested on its predictions using new data, not whether it conforms to the data that gave rise to it --- this is the \gls{verification.stage}\index{science!verification stage} of the scientific method. Hypotheses that run this gauntlet are tentatively accepted by the scientific community; those that do not are discarded (in principle at least).

At this writing, econophysics is still in the observational/experimental stage, with few theories yet formulated. But the research activity is intense, especially for \gls{high-frequency}\index{high-frequency} trading, and one should expect it to enter the hypothesis forming stage in the future. 

\subsection{Market Personification}\label{OACTOM:SS:MarketPersonification}\index{philosophical approaches!market personification|textbf}\index{market personification|textbf}

Na\"{i}ve investors have a tendency to endow a market with human emotions, as if through caprice it were at times nice, at others angry: ``Despite recent choppiness, an optimistic market finished up an agreeable one half percent.'' Not unlike the shenanigans of Greek Gods whose motives, we are told, lie hopelessly beyond the comprehension of mere mortals. 

Such artistic license may be harmless, but its cousin, the ``they'' of market lore, is not. ``They took the market down today.'' ``They really liked XYZ's earnings!'' Though superficial, these sentiments are too persistent to be entirely ignored. Their unstated premise, obviously, is that the market's puppet-strings are manipulated by powerful, if unseen, actors. And investing would then become divination or the fastidious parsing of pronouncements from acknowledged puppet-masters (which seems remarkably similar to today's ``FED watching''). Yet surprisingly, a version of this idea does make sense. In the spirit of our scientific/statistical/gambling approach, it is based not on whimsy or divination, but on an analytical analysis that is presented in the books \cite{moffitt2017V1,moffitt2017V2}.

\subsection{The Strategic Approach}\label{OACTOM:SS:TheStrategicApproach}\index{philosophical approaches!strategic|textbf}\index{strategic approach|textbf}

The strategic approach is essentially a practical, non-ideological one that draws its inspiration from gambling. Professional gamblers view the market as a game\index{games \& game theory} with a large number of players --- they realize that some players are more informed, more skilled than others. They realize that some are very poor players. They understand that emotional people usually don't make good decisions. They realize that some markets are nearly zero-sum\index{games \& game theory!zero-sum}, i.e. futures and options markets and some are not, i.e. equity markets. In an analysis of equity markets, gamblers realize that the players includes corporate management, not just investors and speculators.

There are really only two principles in gambling theory: (1) never bet unless you have an \gls{edge}\index{edge}\index{edge} (= positive expected return) and (2) bet an amount appropriate for that edge. Most long-term losers at games\index{games \& game theory} of chance violate one of these two principles.

Perhaps surprising to economists, most \glslink{professional.gamblers}{professional gamblers}\index{professional gamblers}\index{gambler, professional} are pretty much risk neutral (\cite{RePEc:nbr:nberwo:12767}). For them, each opportunity to bet is viewed as an opportunity to maximize expected return; gamblers prefer many small bets in order to avoid huge swings in capital. Therefore, professional gamblers make bets in some proportion to their \gls{bankroll}\index{bankroll} (investment funds). If they have a small bankroll, they make small bets; if they have a large bankroll, they make large bets. Professional gamblers view their work as one long game\index{games \& game theory}. Odds will change, games will change, competitors will change ---  but it's still just one long game. The secret is to be tolerant to boredom, since the activity consists of using the same methods with mind-numbing, ultra-consistency over years and years.\footnote{A quote from \gls{Soros.George}\index{Soros, George}, ``If investing is entertaining, if you're having fun, you're probably not making any money. Good investing is boring.''} 

Not surprisingly, \glslink{professional.gamblers}{professional gamblers}\index{professional gamblers}\index{gambler, professional} are usually good money managers. They know that their bankroll is all they've got to keep them in the game. Incidentally, the reason professional gamblers are largely risk-neutral is that being risk-averse ``gives up edge'', while being risk-seeking needlessly increases risk.
There is a difference between most gambling games\index{games \& game theory} and the market. Most gambling games, e.g. poker, involve only a few players. The same game is played repeatedly. In single sessions, the game involves the same players.  The best \glslink{professional.gamblers}{professional gamblers}\index{professional gamblers}\index{gambler, professional} are good at \gls{reading.people}\index{reading people}, that is, at guessing what they're trying to do. Some behaviors give away what a player is doing --- these are called \glspl{tell}\index{tell}. Interestingly, lots of gamblers have the same \glspl{tell}\index{tell}! For this reason, gamblers find it plausible that market participants also have \glspl{tell}\index{tell}.

So, what does all this gambling stuff have to do with markets? A great deal, actually!

The big difference between the idealistic and the scientific approaches compared to the strategic approach is that the strategic approach explicitly views markets as games\index{games \& game theory} in which some people use inferior strategies. The strategic approach explicitly rejects efficient market ideology as a way of understanding how markets operate. It replaces this with the view that strategic interaction leads to price formation, and that the key to understanding markets is understanding how competing strategies lead to prices.  Efficient markets are a byproduct inasmuch as simple trading patterns are so easily exploited by simple counter-strategies, that they don't occur in easily recognizable forms.

The point of the market game\index{games \& game theory} is to develop strategies that win. Strategies that are revealed, that is, which can be anticipated --- are candidate inferior strategies. 

\section{Examples of a Strategic/Gambling Approach to Markets}\label{OACTOM:S:ExamplesOfAStrategic/GamblingApproachToMarkets}

We offer several examples of a gambling approach to markets, the central topic of the two volume series ``The Strategic Analysis of Financial Markets,'' \cite{moffitt2017V1,moffitt2017V2}.

One rather obvious requirement for getting a gambling edge is choosing a favorable game. A moment's thought suggests some criteria that should guide game selection. First, good gamblers should choose games with weak players. But what constitutes weakness?\footnote{There's an old poker saying: ``Look around the table. If you don't see a sucker, it's you.''} A few common examples: (1) a (relatively) large group that telegraphs the strategies they'll play, (2) a group of players who act on ``automatic pilot,'' using predictable heuristics that don't engage reflective cognition, and (3) players (not necessarily traders or investors) who are constrained to act in a patterned manner, e.g. mutual funds can't short stock, the FED is mandated to discuss policy at scheduled meetings, etc. Second, the ``weak'' players should be sufficiently influential to move markets --- not necessarily the entire market --- but at least a part of it (a stock for instance). Some examples: (1) investors in IRAs on tax day, (2) decisions by the Federal Reserve on announcement days, and (3) corporate management deciding to raise a dividend. Third, markets sometimes exhibit biases that are important for trading. In bull markets, for example, it is usual that bad news is ignored and good news has a (diminishing) positive impact. 

Here are three examples of systems that were developed using the strategic/gambling approach.
\begin{description}
\item{\bf The Tax Day Trade}

The equity curve of Figure \ref{TSAOMM:G:IRSTaxDayStrategy} is discussed thoroughly in \cite{moffitt:TaxDayTrade}. The trade sets up because tax payers procrastinate in deciding to establish IRAs and flood brokers with money on the last day (the day taxes are due) to open one. Brokers place those orders on the next day, causing the S\&P 500 to gain about 1/2\% on average.
 \begin{figure}[h!]
  \centering
  \includegraphics[scale=0.5]{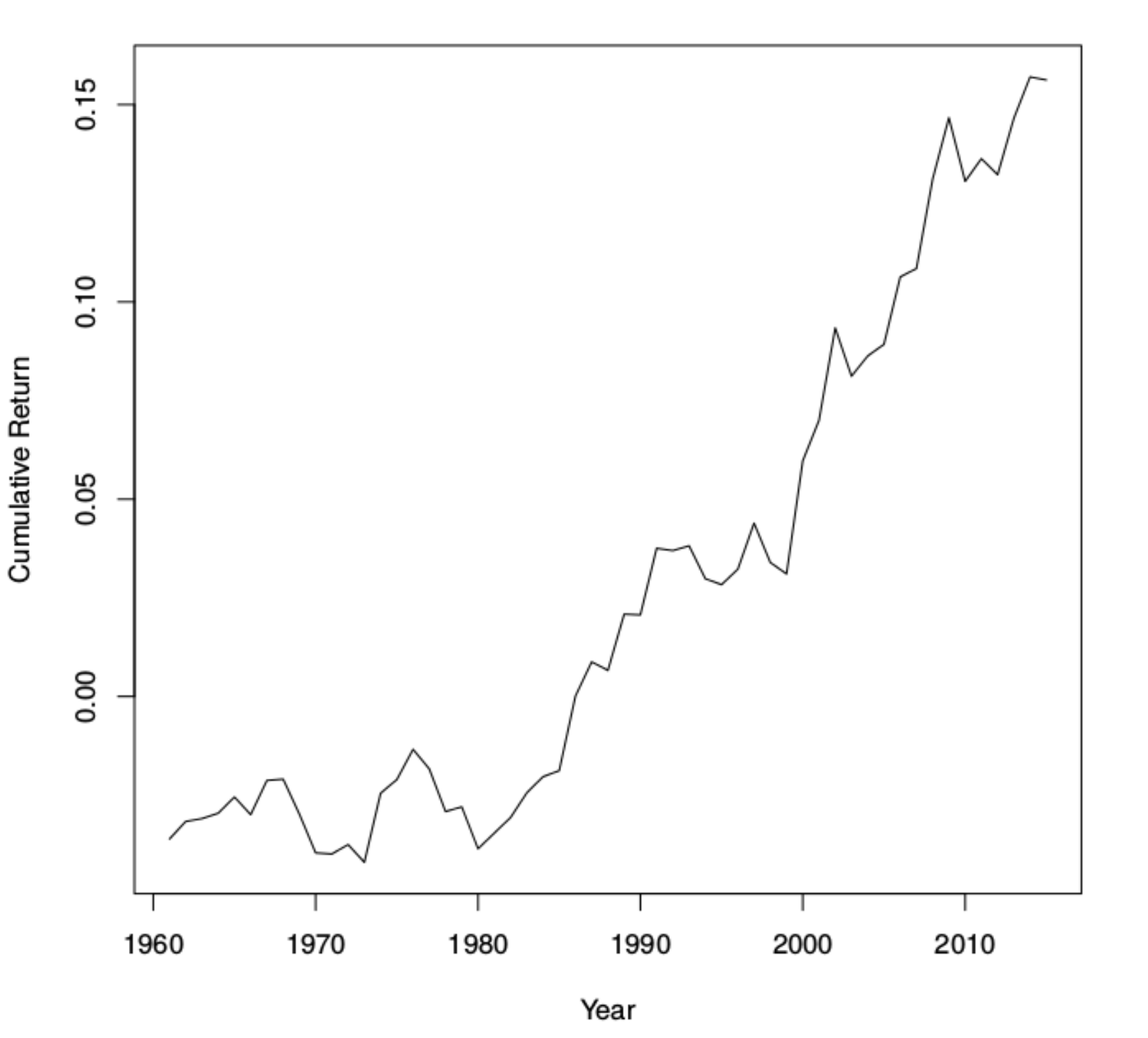}
  \caption{\small{Cumulative returns from buying the S\&P 500 index on the close of U.S. tax day and selling on the close one day later. Commissions and slippage not included.}}
  \label{TSAOMM:G:IRSTaxDayStrategy}
\end{figure}

\item{\bf The Holiday Effect}

The holiday effect, a strategy of buying stocks before a holiday and selling them shortly after the holiday, has been known for a long time, e.g. \cite{fosback1976stock,hirsch1987don,10.2307/2962097}. Older studies were done mostly in the U.S. markets, but the effect exists in many countries (\cite{keim2000security,RePEc:eee:ejores:v:74:y:1994:i:2:p:198-229}).  \cite{ziemba2012calendar} presents an analysis that shows that for U.S. markets in recent years, it has been best to buy on Day -3 and Sell on Day +2.\footnote{Dates are the number of trading days relative to the holiday. Note: In the U.S., markets are closed on all the holidays,}

There have always been two explanations for the holiday effects. The rational explanation is that a few traders sell pre-holiday in order to reduce risk, thus exerting downward pressure which is subsequently corrected post-holiday. The ``animal spirits'' (irrational) explanation is that the warm glow of post-holiday cheer leads to optimistic trading. Since modern finance is dominated by rational theory, it is hardly surprising that the rational explanation holds sway at present.

But a recent article (\cite{9049309720131015}) uses a ``natural experiment'' to reexamine this explanation. The authors note that the Portuguese stock market was ``synchonized'' in 2003 to the European holiday calendar, so that the Portuguese market no longer ceases trading on most Portuguese national holidays, as it previously did. Thus the natural experiment: compare market action on traded (since 2003) Portuguese national holidays to that before 2003. And \ldots surprise, surprise, the price action around Portuguese national holidays is much as it was formerly! Thus gamblers accept holiday anomalies as due mainly to animal spirits, and perform analysis to determine the trading strategies.\footnote{The rational explanation may play a role, but that seems unlikely. After all, if one empirically observes excess pre-holiday returns, wouldn't it be rational to \emph{buy}, not \emph{sell} before a holiday?}

\newpage
\item{\bf A Risk-Aversion Trade}

We won't describe the details of this trade because it would likely cause it to cease working, but here is a little information about it. It is based on investor's risk aversion; its cumulative returns (not compounded) for a seventeen year period after fees and slippage is shown in Figure \ref{OACTM:G:RiskAverseSystem}. The system was designed using strategic principles recommended in the companion paper \cite{moffitt:SSRN:WhyMarketsAreInefficient:2017}. 
\begin{figure}[ht]
  \centering
    \includegraphics[scale=1.00]{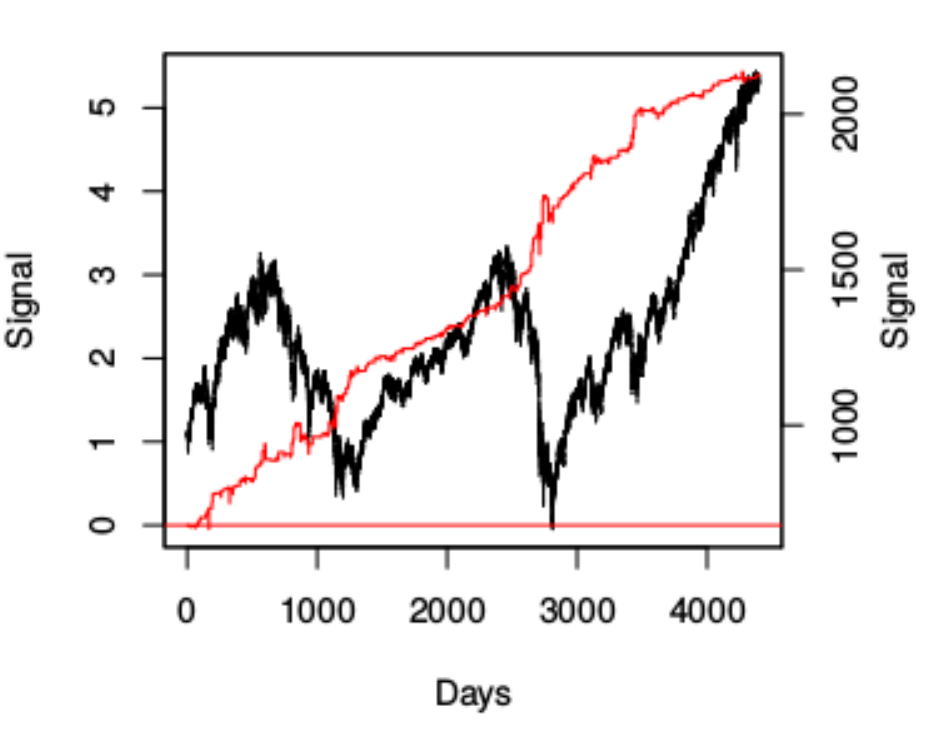}
    \caption{\small{Seventeen years of daily cumulative returns (no compounding) for a trading system based on risk aversion. It averages 25\% per year and has a single drawdown of about 35\% just prior to day 3000. In the portion of the curve from day 3000 to the end, the worst drawdown was about 12\%.}}
     \label{OACTM:G:RiskAverseSystem}
\end{figure}
\end{description}

% with additional discussion of them from \glslink{market.fundamentalism}{market fundamental}\index{philosophical approaches!idealistic}\index{market fundamentalism} and scientific perspectives.\index{philosophical approaches!scientific}\index{scientific approach}
For completeness, we discuss these three examples from the perspective of \gls{market.fundamentalism}\index{philosophical approaches!idealistic}\index{market fundamentalism} and the scientific approach.\index{philosophical approaches!scientific}\index{scientific approach} For each of these three examples, market fundamentalism would argue that either (1) there is a problem with study design, so that the purported effects are spurious, or (2) that the effects are too small to be important to market efficiency or (3) they will be \gls{arbbed.out}\index{arbitrage} soon after becoming known. Since traders \emph{have made (and continue to make) money) with calendar trading} (\cite{ziemba2012calendar}) and since these effects have been around for over a hundred years (\cite{10.2307/2962097}), the only viable explanation is (2). But (2) is problematic because it admits that traders are irrational!

The scientific view of behavioral finance is that known heuristics account for the first two ``anomalies'' but not the third. Neither behavioral finance nor economics considers risk aversion ``irrational.'' Yet it is  --- it gives away money!

\section{Summary}\label{OACTOM:S:Summary}

This article and its companion \cite{moffitt:SSRN:WhyMarketsAreInefficient:2017} dispense with time-worn clich\'{e}s\footnote{Has ``time-worn clich\'{e}'' itself become a time-worn clich\'{e}?} about \glspl{rational.actor}\index{rationality} and efficient markets, which have little to do with how actual market participants behave --- arguing instead that actors are neither rational nor markets efficient\index{efficient market hypothesis (EMH)}. It takes the view that markets are best understood as the result of strategic interaction among the participants. In this view, each strategy can have its day in the sun depending, of course, on the mix of other strategies it faces. Trading based on fundamental analysis, technical analysis and \gls{arbitrage}\index{arbitrage} will be successful, albeit not all the time nor all at the same time. Market movement occurs only when the transacting parties use different strategies; markets respond both to ``external'' news and to ``internal'' effects, that is, to secondary effects due only to strategic interaction, not news.

One cannot gain a durable edge in investing without understanding other participants' strategies, and a critical component of such understanding is investor psychology. Thus the new field of behavioral finance is essential for anyone who wishes to invest or speculate successfully. But one must also understand the basic principles of game analysis and gambling theory, since most patterned phenomena in markets have causes both behavioral and strategic. 

A experienced trader's response to this article might be, ``Isn't this obvious, that good gambling is essential to successful trading?'' My answer: ``Yes, it is obvious.'' But certain of its consequences are \emph{definitely not obvious}. For example, how can you guess --- \emph{without any data analysis} --- that a hypothesized anomaly is like to persist? The obvious answer according to efficient market thinking, that it will disappear after discovery, is absolutely wrong!\footnote{A colleague communicated that Merton Miller told him that revealed anomalies wouldn't disappear immediately, as efficient market theory requires, but would likely take 3 years or so. But this is demonstrably wrong, as the case of market momentum demonstrates.} Or, how can you guess that a bubble is forming, again \emph{without any data analysis}? These are consequences of a strategic analysis, and as I said, they are not obvious.
 
It's time for a revision of market theory. The overwhelming evidence is that markets have never been efficient in the sense of the efficient market hypothesis. Astute readers will realize that the ghosts of bygone inefficiencies still haunt us --- only their shapes have shifted.

\newpage 
%\printglossaries
\bibliographystyle{apalike}
{\footnotesize 
\bibliography{OACTOM.bib}}
\printindex

\end{document}